\documentclass[journal]{IEEEtran}
\IEEEoverridecommandlockouts
\usepackage{url}
\usepackage{cite}
\usepackage{amsmath,amssymb,amsfonts}
\usepackage{graphicx}
\usepackage{textcomp}
\usepackage[utf8]{inputenc}
\usepackage{amssymb}
\usepackage{color}
\usepackage{relsize}
\usepackage{threeparttable}
\usepackage[nolist]{acronym}
\usepackage{multirow}
\usepackage{framed}
\usepackage{soul}
\usepackage{algpseudocode}
\usepackage{algorithm}
\usepackage{amsmath}
\usepackage{amsthm}
\usepackage[squaren,Gray]{SIunits}
\usepackage[dvipsnames]{xcolor}
\usepackage{balance}
\usepackage{comment}
\usepackage{longtable}
\usepackage{graphicx}
\usepackage{afterpage}

\begin{document}

\title{Electromagnetic Signal and Information Theory \\ \vspace{0.20cm} \normalsize Electromagnetically Consistent Communication Models for the Transmission and Processing of Information }
\author{Marco Di Renzo,~\IEEEmembership{Fellow,~IEEE}, and Marco Donald Migliore,~\IEEEmembership{Senior Member,~IEEE \vspace{-0.87cm}}
\thanks{Manuscript received Oct. 21, 2023; revised Dec. 31, 2023.}
\thanks{M. Di Renzo is with Universit\'e Paris-Saclay, CNRS, CentraleSup\'elec, Laboratoire des Signaux et Syst\`emes, 3 Rue Joliot-Curie, 91192 Gif-sur-Yvette, France. (marco.di-renzo@universite-paris-saclay.fr).}
\thanks{M. D. Migliore is with the DIEI, University of Cassino and Lazio Meridionale, Italy. (mdmiglio@unicas.it).}
\thanks{The work of M. Di Renzo was supported in part by the European Commission through the H2020 ARIADNE project under grant agreement number 871464 and through the H2020 RISE-6G project under grant agreement number 101017011, and by the Agence Nationale de la Recherche (France 2030, ANR PEPR Future Networks, grant NF-Founds, 22-PEFT-0010).}
}

\maketitle

\begin{abstract}
In this paper, we present electromagnetic signal and information theory (ESIT). ESIT is an interdisciplinary scientific discipline, which amalgamates electromagnetic theory, signal processing theory, and information theory. ESIT is aimed at studying and designing physically consistent communication schemes for the transmission and processing of information in communication networks. In simple terms, ESIT can be defined as physics-aware information theory and signal processing for communications. We consider three relevant problems in contemporary communication theory, and we show how they can be tackled under the lenses of ESIT. Specifically, we focus on (i) the theoretical and practical motivations behind antenna designs based on subwavelength radiating elements and interdistances; (ii) the modeling and role played by the electromagnetic mutual coupling, and the appropriateness of multiport network theory for modeling it; and (iii) the analytical tools for unveiling the performance limits and realizing spatial multiplexing in near field, line-of-sight, channels. To exemplify the role played by ESIT and the need for electromagnetic consistency, we consider case studies related to reconfigurable intelligent surfaces and holographic surfaces, and we highlight the inconsistencies of widely utilized communication models, as opposed to communication models that originate from first electromagnetic principles.
\end{abstract}
	

\section{Introduction}
{\textcolor{black}{\textbf{Towards sustainable 6G networks} -- The radio communication division of the international telecommunication union (ITU-R) has recently drafted new recommendations for the international mobile telecommunication 2030 (IMT-2030) framework, which is referred to as the sixth generation (6G) of telecommunication standards. In the past decade, several advanced wireless technologies, including small cells, millimeter-wave communications, and massive multiple-input multiple-output (MIMO) systems, have been proposed to enhance the network capacity and to enable ubiquitous wireless connectivity. The practical implementation and deployment of these technologies is, however, often limited by the associated prohibitive energy consumption and expensive hardware equipment. As a result, it has become apparent that 6G communication networks need to undergo a fundamental shift of design paradigm, which requires to include aspects of (energy) sustainability, besides those of  network capacity and connectivity, at the design stage. This change of design paradigm requires radically new physical layer technologies.}}

\textbf{Wave domain information processing} -- {\textcolor{black}{In this context, we are assisting to the upsurge in brand-new technologies for the physical layer, which rely on encoding, processing, and decoding information in the wave domain, i.e., at the electromagnetic level, as opposed to conventional physical layer technologies that rely on digital information processing \cite{an2023stacked}. The advantages of wave domain information processing include improved computational efficiency, simplified hardware architectures, and reduced energy consumption \cite[Table I]{an2023stacked}. This emerging trend has been facilitated by recent results in the field of configurable antennas and, especially, metasurfaces, which are engineered materials capable of processing the electromagnetic waves in the wave domain without the need of analog-to-digital and digital-to-analog conversions \cite{RenzoZDAYRT20}.}}

\textbf{RIS, HoloS, and the like} -- Examples of emerging technologies include (i) spatial, index, media-based, metasurface modulation, which encode information onto physical characteristics of antennas and metasurfaces \cite{LiWR21}; (ii) reconfigurable intelligent surfaces (RISs), which improve the transmission of data by appropriately shaping the propagation of electromagnetic waves at the electromagnetic level, turning radio propagation environments into smart radio propagation environments \cite{RenzoZDAYRT20}; (iii) holographic surfaces (HoloSs), which are continuous-aperture hybrid MIMO systems, where the data encoding and decoding is performed in the wave domain \cite{an2023stacked}; and (iv) stacked intelligent surfaces (SIMs), which are multi-layer metasurface-based devices, which resemble deep neural networks, where the data encoding and encoding is realized through signal processing operations in the wave domain \cite{an2023stacked}.

\textbf{Electromagnetically consistent communications} -- Despite the potential performance gains and applications that these technologies may provide in future wireless networks \cite{SihlbomPR23}, the major limiting factor preventing information, communication, and signal processing theorists from realizing their full potential and unveiling their ultimate performance limits lies in understanding the electromagnetic and physical properties and limitations underpinning them. {\textcolor{black}{Key open problems include how to appropriately model the physics of signal propagation and the processing of signals performed by these emerging devices in the wave domain \cite{RenzoDT22}}}. To overcome this status quo, it is necessary to cut across the current and established disciplinary boundaries between information, signal, and electromagnetic theories. This is the objective of \textit{\textbf{electromagnetic signal and information theory (ESIT)}}.

\textbf{ESIT} -- ESIT is an interdisciplinary scientific discipline, which amalgamates electromagnetic theory, signal processing theory, and information theory, for studying and designing physically consistent (or Maxwellian) communication schemes for the transmission and processing of information in communication networks. In simple terms, \textbf{\textit{ESIT can be defined as physics-aware information theory and signal processing for communications}} -- A concept that is rooted in Dennis Gabor's vision, according to which communication is a branch of physics \cite{1188558}. It is worth mentioning that ESIT or EIT (electromagnetic information theory) is not a new concept. Indeed, the need of reconciling Maxwell's equations with communication theory, and the interrelation between electromagnetic theory and communication theory have been the subject of vivid discussions, e.g., \cite{4685903, 4020419, 4636839}, and the term EIT is well-known in the electromagnetic and antenna communities. For example, we can quote the first paragraph of \cite{4685903}: ``\textit{This research is concerned with the formulation of fundamental wireless communication and antenna engineering problems at the crossroads of the well-established fields of electromagnetic theory and information theory. The interdisciplinary field constituted by such wave- and information-theoretic problems and their solutions can be descriptively termed electromagnetic information theory or, within the antenna focus, antenna information theory. Work in this area, combining wave physics with information theory, has a long history, dating back to the origins of information theory and, in particular, to the pioneering work on light and information by Gabor \ldots}''.

\textbf{Communications: Mathematics or physics?} -- Despite the relevance and the attempts made to reunite information and electromagnetic theories, these disciplines evolved utilizing related but often inconsistent models. Specifically, the design of past and current generations of communication systems has been driven by Shannon's theory \cite{6773024}, which is a \textit{mathematical theory} of communication. The notion of channel capacity as well as the methods, algorithms, and protocols to achieve it are fundamental questions that have driven the design of communication systems and will continue to do so. In that regard, Gabor commented, already in 1953 (five years after the publication of Shannon's mathematical theory), the following \cite{1188558}: ``\textit{Communication theory has up to now developed mainly on mathematical lines, taking for granted the physical significance of the quantities which figure in its formalism. But communication is the transmission of physical effects from one system to another, hence communication theory should be considered as a branch of physics. Thus, it is necessary to embody in its foundations such fundamental physical data}''.

\textbf{{\textcolor{black}{Electromagnetically consistent mathematical abstractions}}} -- Based on Gabor's vision, communication and information are inherently physical phenomena. In current information and communication frameworks, however, the physics of wave propagation is usually abstracted, by treating the generation, transmission, propagation, and manipulation of the electromagnetic waves as mathematical operators, according to Shannon's theory. Although mathematical abstractions and approximations are necessary to design advanced and complex communication systems and to gain so-called ``engineering insights'', in this abstraction and approximation process the electromagnetic consistency of the obtained communication models and signal processing algorithms is often lost. ESIT is aimed at ensuring that Shannon's communication theory is built upon models that are electromagnetically consistent, so as to identify the correct ultimate information theoretic limits of communication systems, and to design the corresponding signal processing algorithms for attaining those limits.

\textbf{Amalgamating Shannon's and Gabor's theories} -- Therefore, the challenge lies in amalgamating Shannon's and Gabor's theories into a cohesive framework, wherein the electromagnetic constraints imposed by Gabor's physics-based theory find inclusion into Shannon's mathematical-based theory of communication. From a physical perspective, the transmission of information requires encoding data into distinguishable states of a physical observable. In wireless communications, the observable is the configuration of the electromagnetic field at a receiving antenna, and a unit of information (a ``bit'') is represented by two different states that are distinguishable in the presence of noise and interference. This approach inherently leads to a ``physical approach'' to information theory, as put forth by Gabor \cite{1188558}. A purely deterministic ``physical theory of communication'' encounters, however, some limitations, since it fails to account for a fundamental characteristic of communication channels: The statistical nature of the transmission of information due to, e.g., the presence of noise, which is considered in Shannon's ``mathematical theory of communication'' \cite{6773024}. The amalgamation of Shannon's and Gabor's theories into a unified framework is the core tenet of ESIT. Embracing the celebrated open systems interconnection model, which is based on abstraction layers, a plausible approach to this end suggests to introduce an additional layer below the physical layer, whose functionality is to map the ``logical'' bits of Shannon's theory into the ``physical'' bits that are associated with different states of the observable \cite{9438650}.

{\textcolor{black}{\textbf{Aim of this paper} -- In this paper, we embrace an electromagnetically consistent approach to communication design, and show how, moving from first electromagnetic principles, we can understand the fundamental performance limits and design principles of emerging physical layer technologies. For example, RISs and HoloSs have been extensively studied \cite{RenzoZDAYRT20}, \cite{RenzoDT22}, but fundamental questions about their modeling and design are still under scrutiny by the communication, metamaterial, and antenna communities, \cite{abs-2308-16856}, \cite{10042168}, \cite{li2023tunable}. Under the lenses of ESIT, we attempt to answer three core and open questions}}.
\begin{enumerate}
\item \textbf{{\textcolor{black}{Subwavelength}} design} -- A metasurface is usually defined as an engineered surface whose radiating (scattering) elements have sizes and interdistances smaller than half of the wavelength, in order to realize sophisticated wave transformations at high power efficiency. {\textcolor{black}{Anomalous reflection is the simplest and canonical wave transformation in RIS-assisted communications \cite{RenzoDT22}}}. According to antenna theory and the sampling theorem, metasurface designs based on half-wavelength scattering elements and interdistances are \textit{de facto} considered sufficient and are often assumed optimal in the very back of communication theorists' minds. \textit{Are there and what are the theoretical and practical (implementation) needs or advantages of subwavelength metasurfaces?}
\item \textbf{Electromagnetic mutual coupling} -- Assume that there exist either advantages from the theoretical standpoint or needs from the practical standpoints for using subwavelength metasurfaces. Thus, the electromagnetic mutual coupling among the radiating (scattering) elements cannot be overlooked anymore \cite{RenzoDT22}. {\textcolor{black}{\textit{What electromagnetically consistent methods to utilize for modeling the mutual coupling? What signal processing methods to leverage for optimizing a metasurface by taking the mutual coupling into account at the design stage?}}}
\item \textbf{Spatial multiplexing in the near field} -- To be beneficial, RISs and HoloSs need to be electrically large, with sizes of the order of {\textcolor{black}{ten-hundreds}} wavelengths \cite{SihlbomPR23}. {\textcolor{black}{In many communication scenarios, this results in transmission distances for which two metasurfaces are located in the (radiating) near field of one another \cite[Fig. 4]{SihlbomPR23}. Far field propagation is, however, the \textit{de facto} model adopted to design current wireless systems. Thus, key questions become pertinent: \textit{What electromagnetically consistent models to utilize for near-field communications?} \textit{What are the fundamental performance limits of near-field communications, e.g., in terms of spatial multiplexing gain (degrees of freedom), and optimal communication (encoding and decoding) waveforms? Also, what electromagnetically consistent mathematical tools to use?}}}
\end{enumerate}

\textbf{Paper organization} -- Next, we elaborate on these questions, {\textcolor{black}{by staying grounded in first electromagnetic principles and providing concrete examples available in the literature.}}

\section{Subwavelength Design}
In this section, we elaborate on subwavelength designs that are considered a distinctive feature of metasurfaces, as opposed to conventional antenna arrays \cite{RenzoDT22, 10042168}. To understand the theoretical and practical needs for utilizing  scattering elements whose sizes and interdistances are smaller than $\lambda/2$, with $\lambda$ being the wavelength of the electromagnetic signal, we invoke the plane wave spectrum representation of an electromagnetic field \cite{1140193} and Petersen-Middleton's theorem, i.e., the sampling theorem in $N$-dimensional Euclidean spaces \cite{PETERSEN1962279}, which is a generalization of Nyquist-Shannon's sampling theorem in one-dimensional Euclidean spaces (i.e., for time-domain signals).

\textbf{Plane wave spectrum representation} -- We consider a monochromatic electromagnetic field whose electric field evaluated at the observation point $\left( {x,y,z} \right)$ is denoted by ${\bf{E}}\left( {x,y,z} \right)$.  For simplicity, the time-dependence ${e^{j\omega t}}$ is omitted where $j = \sqrt{-1}$ is the imaginary unit. For example, ${\bf{E}}\left( {x,y,z} \right)$ may be the electric field scattered by an RIS that is located at $z=0$ and is parallel to the $xy$-plane. In Cartesian components, ${\bf{E}}\left( {x,y,z} \right) = {E_x}\left( {x,y,z} \right){\bf{\hat x}} + {E_y}\left( {x,y,z} \right){\bf{\hat y}} + {E_z}\left( {x,y,z} \right){\bf{\hat z}}$.

The plane wave spectrum representation allows us to express the scattered field ${\bf{E}}\left( {x,y,z} \right)$ as an integral of plane waves. Specifically, ${\bf{E}}\left( {x,y,z} \right)$ evaluated in the $z \ge 0$ half-space can be formulated as follows \cite[Eq. (19.5.8)]{Orfanidis}, {\textcolor{black}{\cite{9798854}}:
\begin{equation} \label{Eq:PlaneWaveSpectrum_E}
{\bf{E}}\left( {x,y,z} \right) = \int \int {{{\bf{\mathord{\buildrel{\lower3pt\hbox{$\scriptscriptstyle\frown$}}
\over E} }}}}\left( {{\kappa _x},{\kappa _y}} \right) {e^{ - j{\kappa _x}x}}{e^{ - j{\kappa _y}y}}{e^{ - j{\kappa _z}z}}\frac{{d{\kappa _x}d{\kappa _y}}}{{{{\left( {2\pi } \right)}^2}}} \nonumber
\end{equation}
where $\int {\int {\left(  \cdot  \right)} }  = \int_{ - \infty }^{ + \infty } {\int_{ - \infty }^{ + \infty } {\left(  \cdot  \right)} }$ and, to ensure that Maxwell's equations are fulfilled, the following relations need to hold:
\begin{equation} \label{Eq:FourierTransform_E}
{{{\bf{\mathord{\buildrel{\lower3pt\hbox{$\scriptscriptstyle\frown$}}
\over E} }}}_{\tan }}\left( {{\kappa _x},{\kappa _y}} \right)  = \hspace{-0.2cm} \int \int {{{\bf{E}}_{\tan }}\left( {{x^{\prime}},{y^{\prime}},{z^{\prime}} = {0^ + }} \right){e^{ j{\kappa _x}{x^{\prime}}}}{e^{ j{\kappa _y}{y^{\prime}}}}d{x^{\prime}}d{y^{\prime}}} \nonumber
\end{equation}
where ${{{\bf{\mathord{\buildrel{\lower3pt\hbox{$\scriptscriptstyle\frown$}}
\over E} }}}_{\tan }}\left( {{\kappa _x},{\kappa _y}} \right)  = {{\mathord{\buildrel{\lower3pt\hbox{$\scriptscriptstyle\frown$}}
\over E} }_x}\left( {{\kappa _x},{\kappa _y}} \right){\bf{\hat x}} + {{\mathord{\buildrel{\lower3pt\hbox{$\scriptscriptstyle\frown$}}
\over E} }_y}\left( {{\kappa _x},{\kappa _y}} \right){\bf{\hat y}}$, ${{\bf{E}}_{\tan }}\left( {x,y,z} \right) = {E_x}\left( {x,y,z} \right){\bf{\hat x}} + {E_y}\left( {x,y,z} \right){\bf{\hat y}}$, and
\begin{equation} \label{Eq:FourierEZ}
{{\mathord{\buildrel{\lower3pt\hbox{$\scriptscriptstyle\frown$}}
\over E} }_z}\left( {{\kappa _x},{\kappa _y}} \right) =  - \frac{{{\kappa _x}{{\mathord{\buildrel{\lower3pt\hbox{$\scriptscriptstyle\frown$}}
\over E} }_x}\left( {{\kappa _x},{\kappa _y}} \right) + {\kappa _y}{{\mathord{\buildrel{\lower3pt\hbox{$\scriptscriptstyle\frown$}}
\over E} }_y}\left( {{\kappa _x},{\kappa _y}} \right)}}{{{{\kappa _z}\left( {{\kappa _x},{\kappa _y}} \right)}}} \nonumber
\end{equation}
with $\kappa  = 2\pi/\lambda$ and ${\kappa _z}\left( {{\kappa _x},{\kappa _y}} \right)$ is given by
\begin{align}
& {\kappa _z}\left( {{\kappa _x},{\kappa _y}} \right) = \sqrt {{\kappa ^2} - \kappa _x^2 - \kappa _y^2} \, \, \, \, \quad \quad \, \, {\rm{if}}\quad \kappa _x^2 + \kappa _y^2 \le {\kappa ^2} \label{Eq:realKappaZ} \\
& {\kappa _z}\left( {{\kappa _x},{\kappa _y}} \right) = -j\sqrt { - {\kappa ^2} + \kappa _x^2 + \kappa _y^2} \quad {\rm{if}}\quad \kappa _x^2 + \kappa _y^2 > {\kappa ^2} \label{Eq:imagKappaZ}
\end{align}
Therefore, any electric field ${\bf{E}}\left( {x,y,z} \right)$ can be expressed as a function of the electric field ${\bf{E}}_{\tan }\left( {{x^{\prime}},{y^{\prime}},{z^{\prime}} = {0^{+}}} \right)$ that is evaluated on the plane $z=0$. The reference plane $z=0$ is chosen with no loss of generality. The spectrum of plane waves that constitutes the electric field can be split in two groups.
\begin{itemize}
\item \textbf{Propagating waves} -- The plane waves with $\kappa_z$ given by \eqref{Eq:realKappaZ} are referred to as propagating waves. The inequality in \eqref{Eq:realKappaZ} defines the so-called ``visible range''. The plane waves of the spectrum representation that lie inside the visible range contribute to the electric field in the far field.
\item \textbf{Evanescent waves} -- The plane waves with $\kappa_z$ given by \eqref{Eq:imagKappaZ} are referred to as evanescent waves. The inequality in \eqref{Eq:imagKappaZ} defines the ``invisible range''. The evanescent waves decay exponentially along the ${\bf{\hat z}}$-axis and are negligible far away from the reference plane $z=0$, i.e., for $z \gg \lambda$. {\textcolor{black}{Thus, the evanescent waves do not radiate, but they contribute to the effective design of the radiating/scattering system, by modifying the surface current density distribution. To make the evanescent waves manifest in the field, they need to be properly excited. This is detailed next.}}
\end{itemize}

\textbf{Petersen-Middleton's sampling theorem} -- The decomposition of the plane wave spectrum of an electromagnetic field into propagating and evanescent waves is the key aspect to understand the theoretical motivation behind subwavelength designs. Let us scrutinize \eqref{Eq:realKappaZ} and \eqref{Eq:imagKappaZ}. Two cases are of interest.
\begin{itemize}
\item \textbf{$\lambda/2$-spacing is sufficient} -- Assume that the observation point is located sufficiently far away from the reference plane $z=0$, so that only the propagating waves exist and there are no evanescent waves. From \eqref{Eq:imagKappaZ}, we evince that the spatial bandwidth of the electric field in the wavenumber domain, i.e., the support of the plane wave spectrum representation, is finite. Petersen-Middleton's sampling theorem can, thus, be applied, and the electric field can be sampled (and reconstructed) at points, located on a regular grid on the $x$ and $y$ axis, that are spaced by
\begin{equation}
\Delta x = \Delta y =\Delta s = \pi /\kappa  = \lambda /2 \nonumber
\end{equation}
since $\kappa$ is the largest value in the wavenumber domain, i.e., the admissible plane waves are, according to \eqref{Eq:realKappaZ}, those whose values of $(\kappa_x, \kappa_y)$ lie in the visible range. In a one-dimensional space, e.g., for signals in the time domain, this corresponds to a band limited signal whose sampling period is the reciprocal of the bandwidth, as dictated by Nyquist-Shannon's sampling theorem.
\item \textbf{$\lambda/2$-spacing is not sufficient} -- Assume that the observation point is located sufficiently close to the reference plane $z=0$, so that the evanescent waves cannot be ignored. From \eqref{Eq:imagKappaZ}, we evince that the spatial bandwidth of the electric field in the wavenumber domain is not finite: An infinite number of plane waves exist and Petersen-Middleton's sampling theorem cannot be applied. Assume, however, that the evanescent waves whose amplitude is sufficiently small, by a predetermined amount, can be ignored in practice. Denote by $\kappa_{xm}$ and $\kappa_{ym}$ the magnitude of the largest $\kappa_x$ and $\kappa_y$, respectively, of the last non-negligible evanescent wave. Then, the spatial bandwidth of the electric field in the wavenumber domain can be assumed approximately finite, and Petersen-Middleton's sampling theorem can be applied. Specifically, the electric field can be approximately sampled (and reconstructed) at points that are located on a regular grid on the $x$ and $y$ axis, and that are spaced by
\begin{equation}
\Delta x = \pi /\kappa_{xm}  < \lambda /2, \quad \Delta y = \pi /\kappa_{ym}  < \lambda /2 \label{Eq:Samplingy} \nonumber
\end{equation}
{\textcolor{black}{since $(\kappa_{xm}, \kappa_{ym})$ are the largest values of $(\kappa_x, \kappa_y)$ for which the evanescent waves cannot be ignored. Notably, the sampling spacing approaches zero for large values of $\kappa_{xm}$ and $\kappa_{ym}$, leading to a continuum if all the evanescent waves cannot be ignored. In practice, the plane waves that effectively contribute to the electric field are those whose values of $(\kappa_x, \kappa_y)$ are within the disk of radius $\sqrt{\kappa_{xm}^2 + \kappa_{ym}^2} >\kappa$. The relevance of this result for optimizing, e.g., an RIS, is that its reflection coefficient is defined near the reference plane, where the evanescent waves are not negligible and hence cannot be ignored.}}
\end{itemize}

{\textcolor{black}{\textbf{Evanescent waves for wireless: Why?}}} -- The plane wave spectrum representation of an electric field and Petersen-Middleton's sampling theorem provide the theoretical departing point to understand subwavelength designs for metasurfaces. In wireless communications, however, we are typically interested in evaluating the electric field re-radiated by, e.g., an RIS at observation points where the evanescent waves are absent. In fact, even in the radiating near field, the evanescent waves can usually be ignored in wireless applications. Thus, it needs to be understood  why subwavelength structures are beneficial to design metasurfaces for wireless communications. {\textcolor{black}{Generally speaking, it is possible to calculate a current distribution that radiates any intended far-field electromagnetic wave by solving an inverse source problem \cite{RenzoDT22}. These problems have an infinite number of solutions that provide the same far field, which include solutions that rely on controlling the evanescent waves and that may offer design advantages. To better understand this concept, we present three examples.}}

\textbf{{\textcolor{black}{Example 1:}} Perfect wave control via Bochner’s generalization of Poisson’s summation formula} -- In \cite{HANSEN2021102791}, the author studies the optimal current distribution that needs to be impressed on an infinitely extended metasurface that is illuminated by an external source, in order to {\textcolor{black}{re-radiate}} some predetermined plane waves. This is the core function {\textcolor{black}{of RIS}}. For analytical tractability, the author models the metasurface as an array of line sources. By using Bochner’s generalization of Poisson’s summation formula, the author proves that it is not possible to {\textcolor{black}{re-radiate}} a single plane wave by using a discrete array of line sources, but a continuous source distribution is needed. This is because the design objective, i.e., to launch a single plane wave, implies the requirement of perfectly controlling all the possible {\textcolor{black}{re-radiated}} waves, including the evanescent waves. This example is in agreement with the plane wave spectrum representation and Petersen-Middleton's sampling theorem, and motivates, from the theoretical standpoint, the need of subwavelength designs for controlling, with high accuracy, the radiated electromagnetic waves.

\begin{figure}[!t]
\centering
\includegraphics[width=0.90\columnwidth]{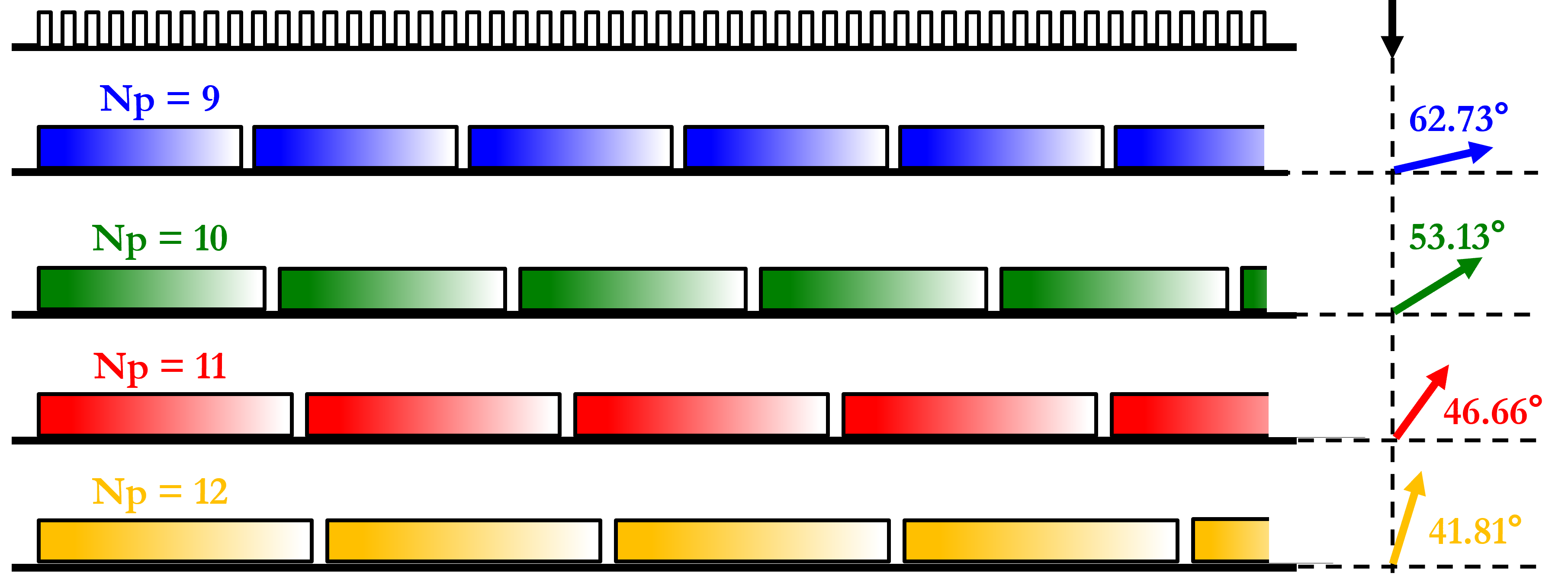}
\caption{{\textcolor{black}{Example of periodic design: The period ($N_p$) of the phase gradient applied by the metasurface determines the angle of reflection. Different colors (including the gradient fill in each period) correspond to different phase shifts. The size of the scattering elements determines the angles of reflection allowed.}}}
\label{Fig:Fig_PeriodicDesign} \vspace{-0.5cm}
\end{figure}
\textbf{{\textcolor{black}{Example 2:}} Perfect reflection via periodic designs} -- There exist several optimal and sub-optimal criteria for the design of metasurfaces that implement specified wave transformations, including anomalous reflection \cite{10042168}. Consider the case study of a plane wave that illuminates an infinitely extended metasurface with an angle of incidence $\theta_i$, and that the metasurface is designed to {\textcolor{black}{re-radiate}} the impinging wave towards an angle of reflection $\theta_r$. The surface impedance (or equivalently the surface reflection coefficient) that characterizes the metasurface is a periodic function with period $D = \lambda/\left| {\sin(\theta_r)-\sin(\theta_i)} \right|$ \cite{RenzoDT22}. The typical criterion utilized to realize this function with a static (non-reconfigurable) metasurface consists of carefully engineering (in size and shape) the scattering elements within one period. The one-period structure is referred to as a supercell, and the metasurface is obtained by repeating several supercells. This approach simplifies the design of metasurfaces, since a single period needs to be optimized, in lieu of the entire metasurface. To ensure a discrete number of scattering elements within a period $D$, their size and spacing $\delta$ cannot be, in general, $\lambda/2$, but need to be smaller than $\lambda/2$, i.e., $\delta \le \lambda/2$. If, e.g., $\theta_i = 0$ and $\theta_r = 70^\circ$, the period is $D \approx 1.064 \lambda$. Hence, scattering elements spaced at $\lambda/2$ cannot provide a periodic metasurface, but this is possible if the ratio $N_p=D/\delta$ is an integer number. {\textcolor{black}{If the metasurface is dynamic (reconfigurable) and is intended to {\textcolor{black}{re-radiate}} the incident electromagnetic wave towards a direction $\theta_r$ that changes over time, the period $D$ needs to be controlled over time, by adjusting the phase response of each scattering element through electronic circuits. This is illustrated in Fig. \ref{Fig:Fig_PeriodicDesign}. In this case, the smaller the size and the spacing among the scattering elements are, the more finely controlled the direction $\theta_r$ is. Infinitesimally small scattering elements are needed to {\textcolor{black}{re-radiate}} the incident signal towards any direction $\theta_r$. In general, the (finite) size of the scattering elements determines the discrete set of directions $\theta_r$ towards which the incident wave can be reflected \cite[Table 1]{RenzoDT22}, as shown in Fig. \ref{Fig:Fig_PeriodicDesign}. The design of periodic metasurfaces is a practical example that motivates subwavelength designs, i.e., the design (optimization) complexity reduces to a single period.}}

\textbf{{\textcolor{black}{Example 3:}} Perfect reflection via surface waves} -- Even though the evanescent waves are not observable in {\textcolor{black}{(and hence do not modify) the far field, they can be utilized for obtaining the desired {\textcolor{black}{re-radiated}} electromagnetic waves in the far field, while fulfilling specified design and implementation requirements.}} Consider again the example of an anomalous reflector, and the case study of a plane wave that needs to be bent from the angle $\theta_i$ towards the angle $\theta_r$. If no electromagnetic fields other than the incident and reflected plane waves are allowed in the reflection half plane, the resulting metasurface is characterized by a surface impedance whose real part needs to take positive and negative values \cite{RenzoDT22}. This corresponds to the need of locally amplifying and attenuating the incident plane wave along the metasurface. In theory, a solution to this problem is to utilize power amplifiers: An option that is not simple to realize in practice, and that results in an active metasurface \cite{10042168}. {\textcolor{black}{Two desirable objectives are, on the other hand, to mimic the power gain-loss optimal behavior without using power amplifiers, thus keeping the metasurface nearly passive, and to minimize the electric losses, as they reduce the power efficiency of the metasurface. This corresponds to the design criterion that the real part of the surface impedance is equal to zero (the surface impedance is purely imaginary). The careful design of surfaces waves, which are evanescent waves, offers a solution to design such a metasurface. In \cite{8358753}, e.g., a set of evanescent waves is optimized to make the metasurface locally lossless, while obtaining almost the same performance as a theoretically optimal metasurface with power losses and gains. The design of metasurfaces that rely on exciting and controlling surface (evanescent) waves requires subwavelength structures, as dictated by the plane wave spectrum representation and Petersen-Middleton's sampling theorem. This approach is useful to design nearly passive metasurfaces with a high power efficiency, even for large angles of deflection, i.e., the difference between $\theta_r$ and $\theta_r$ is greater than $45^\circ-60^\circ$.}}

\textbf{Takeaway} -- {\textcolor{black}{Moving from first electromagnetic principles, we motivated needs and potential benefits of subwavelength designs from the theoretical (plane wave spectrum, Petersen-Middleton's sampling theorem, Bochner’s generalization of Poisson’s summation formula) and practical (periodic designs, perfect reflection with lossless metasurfaces) standpoints. Subwavelength metasurface implementations with no power gains and losses are useful to realize loaded aperiodic gratings as well \cite{li2023tunable}, which are often used in communications \cite{RenzoDT22}.}}

\section{Electromagnetic Mutual Coupling}
Based on Section II, it is necessary to utilize appropriate models for metasurfaces, i.e., electromagnetically consistent, that account for the electromagnetic mutual coupling among the scattering elements, and to develop signal processing algorithms that consider the electromagnetic mutual coupling at the design stage. In this section, we elaborate on multiport network theory \cite{IvrlacN10}, as a candidate approach for modeling and optimizing subwavelength metasurfaces, in which the mutual coupling is leveraged for achieving better performance {\textcolor{black}{\cite{9838533}}}. Also, we show how multiport network theory allows us to understand fundamental aspects of the scattering from engineered surfaces that simplified models are unable to capture.

\textbf{Multiport network theory} -- {\textcolor{black}{In a multiport network model \cite{IvrlacN10}, each scattering element of a (subwavelength) metasurface corresponds to a port loaded by tunable electronic circuits.}} Each scattering element may be connected to an independent electronic circuit or the scattering elements may be connected to the others through additional electronic circuits. The former and simpler design is referred to as single-connected implementation. The latter and more complex design is referred to as beyond diagonal implementation, and it enables a better control of the mutual coupling at the expenses of a higher implementation complexity \cite{li2023diagonal}. A multiport network model is characterized by a matrix that describes the input-output response between every pair of ports. The input-output relation at each pair of ports may be formulated in terms of impedances, which relate the voltages and currents, or in terms of scattering coefficients, which relate the reflected and incident waves. The electromagnetic mutual coupling among the scattering elements is captured by the off-diagonal elements of the impedance and scattering matrices.

{\textcolor{black}{\textbf{Example: RIS-aided communications} -- Consider an RIS-aided system as an example, the end-to-end channel, i.e., the ratio between the voltages at the ports of a multi-antenna receiver and transmitter, can be formulated as follows \cite{DR1}:}}
\begin{equation} \label{Eq:HS}
{{\bf{H}}_{\rm{S}}} = {{\bf{S}}_{{\rm{RT}}}} + {{\bf{S}}_{{\rm{RS}}}}{\left( {{\bf{I}} - {{\bf{\Gamma }}_{\rm{S}}}{\mathbf{S}_{{\rm{emc}}}}} \right)^{ - 1}}{{\bf{\Gamma }}_{\rm{S}}}{{\bf{S}}_{{\rm{ST}}}}
\end{equation}
\begin{equation} \label{Eq:HZ}
{{\bf{H}}_{\rm{Z}}} = \frac{1}{{2{Z_0}}}\left( {{{\bf{Z}}_{{\rm{RT}}}} - {{\bf{Z}}_{{\rm{RS}}}}{{\left( {{\mathbf{Z}_{{\rm{emc}}}} + {{\bf{Z}}_{\rm{S}}}} \right)}^{ - 1}}{{\bf{Z}}_{{\rm{ST}}}}} \right)
\end{equation}
in terms of scattering matrices in \eqref{Eq:HS} and impedance matrices in \eqref{Eq:HZ}. {\textcolor{black}{As detailed in \cite{abs-2308-16856}, ${{\bf{H}}_{\rm{S}}}$ is obtained by assuming that the ports of the transmitter and receiver are matched for zero reflection (``match terminated'')}}, and ${{\bf{H}}_{\rm{Z}}}$ is obtained under the additional assumption of simplifying some terms that are negligible in the far field of each RIS scattering element \cite{DR1}, \cite{abs-2308-16856} (sometimes referred to as unilateral approximation \cite{li2023diagonal}). {\textcolor{black}{The channel models in \eqref{Eq:HS} and \eqref{Eq:HZ} are equivalent, and the relation between them can be found in \cite{li2023diagonal}.}} The meaning of the symbols utilized in \eqref{Eq:HS} and \eqref{Eq:HZ} is the following:
\begin{itemize}
\item $Z_0$ is a reference impedance and usually $Z_0 = 50$ Ohm;
\item ${\bf{I}}$ is the identity matrix of appropriate size;
\item {\textcolor{black}{${{\bf{Z}}_{{\rm{RT}}}}$, ${{\bf{Z}}_{{\rm{ST}}}}$, and ${{\bf{Z}}_{{\rm{RS}}}}$ are the impedance matrices between transmit and receive antennas, transmit antennas and RIS elements, and RIS elements and receive antennas;}}
\item ${{\bf{Z}}_{{\rm{S}}}}$ is the matrix of tunable loads connected to the ports of the RIS; ${{\bf{\Gamma}}_{{\rm{S}}}}$ is the matrix of reflection coefficients at the ports of the RIS. The matrices ${{\bf{\Gamma}}_{{\rm{S}}}}$ and ${{\bf{Z}}_{{\rm{S}}}}$ are {\textcolor{black}{related}} through the identity ${{\bf{\Gamma }}_{\rm{S}}} = {\left( {{{\bf{Z}}_{{\rm{S}}}} + {Z_0}{\bf{I}}} \right)^{ - 1}}\left( {{{\bf{Z}}_{{\rm{S}}}} - {Z_0}{\bf{I}}} \right)$.
\item {\textcolor{black}{The matrices $\mathbf{S}_{{\rm{emc}}}$ and $\mathbf{Z}_{{\rm{emc}}}$ model the electromagnetic mutual coupling. If the mutual coupling is negligible, they are approximately diagonal matrices; otherwise, they are full matrices. $\mathbf{S}_{{\rm{emc}}}$ and $\mathbf{Z}_{{\rm{emc}}}$ are related through the identity ${{\bf{S}}_{{\rm{emc}}}} = {\left( {{{\bf{Z}}_{{\rm{emc}}}} + {Z_0}{\bf{I}}} \right)^{ - 1}}\left( {{{\bf{Z}}_{{\rm{emc}}}} - {Z_0}{\bf{I}}} \right)$.}}
\item {\textcolor{black}{The matrices ${{{\bf{\Gamma }}_{\rm{S}}}}$ and ${{{\bf{Z}}_{\rm{S}}}}$ model the tunable circuits. In single-connected and beyond diagonal RISs, ${{{\bf{\Gamma }}_{\rm{S}}}}$ and ${{{\bf{Z}}_{\rm{S}}}}$ are diagonal and, in general, full matrices, respectively.}}
\end{itemize}
The analytical formulations in \eqref{Eq:HS} and \eqref{Eq:HZ} resemble the channel of a MIMO communication link in free space, as usually considered in communication theory. A multiport network model leads, therefore, to a convenient analytical formulation for performance evaluation and signal processing design and optimization. To utilize the multiport network models in \eqref{Eq:HS} and \eqref{Eq:HZ} is necessary to compute the scattering matrix $\mathbf{S}_{{\rm{emc}}}$ or the impedance matrix $\mathbf{Z}_{{\rm{emc}}}$, which model the mutual coupling.

{\textcolor{black}{\textbf{Computation of $\mathbf{S}_{{\rm{emc}}}$ and $\mathbf{Z}_{{\rm{emc}}}$} -- The matrices $\mathbf{S}_{{\rm{emc}}}$ and  $\mathbf{Z}_{{\rm{emc}}}$ may be obtained through full-wave electromagnetic simulations.}} This approach is often the preferred choice in the electromagnetic community, as it offers accurate estimates for the mutual coupling. Also, realistic scattering elements and tunable circuits may be considered. This approach may not be, however, the most appropriate option in communications and signal processing. In these fields of research, in fact, design parameters like the size, shape, inter-distance, and the load impedances are key variables that are intended to be optimized. If the output from state-of-the-art electromagnetic simulators is parameter-dependent, the applications in communications and signal processing may be limited. As a minimum requirement, the matrices $\mathbf{S}_{{\rm{emc}}}$ and $\mathbf{Z}_{{\rm{emc}}}$ need to be provided for arbitrary tunable loads, and the impact of the tunable loads needs to be known in mathematical terms, in order to consider the electronic circuits as optimization variables.

{\textcolor{black}{\textbf{Modeling the electromagnetic mutual coupling: The role of ESIT} -- Modeling the mutual coupling is a concrete example that unveils the distinct trait of ESIT, when compared with electromagnetic theory (usually simulation-based and electromagnetically consistent) and communication theory (mathematical-based and often electromagnetically inconsistent): ESIT is aimed to develop mutual coupling models that are electromagnetically consistent, are parametric as a function of key optimization variables, and are mathematically tractable. In the context of wireless communications, these are prerequisites for (i) unveiling fundamental design principles and scaling laws; (ii) formulating and solving network optimization problems of practical relevance; and (iii) providing guidelines to develop optimal signal processing algorithms, e.g., encoding, decoding, and channel estimation schemes.}}

\textbf{Loaded thin wire dipole model for metasurfaces} -- {\textcolor{black}{A model for metasurfaces that has these desirable features was introduced in \cite{1236083}. Therein, the authors proved that natural composite materials and electrically controlled engineered materials (metamaterials) can be modeled as arrays of thin wire dipoles loaded by lumped impedances. Using Clausius-Mossotti's formula, the authors have shown that the lumped loads depend on the physical properties (e.g., the permittivity) of the materials and the wave transformations that need to be realized. Thus, a dynamic metasurface can be modeled as an array of subwavelength loaded thin wire dipoles. This model is electromagnetically consistent and mathematically tractable, since the thin wire dipole is one of the few scattering elements for which the current distribution and the scattered field can be formulated in an analytical (but often approximated) form. The authors of \cite{DR1} have moved from the dipole-based model in \cite{1236083} and have derived an end-to-end multiport network model for RIS-aided channels. The model in \cite{1236083} has recently been generalized in \cite{akrout2023physically} for application to Chu's scattering elements. Existing multiport network models for metasurfaces rely, however, on one key assumption \cite{DR1}: The scattering elements are treated as canonical minimum scattering antennas.}}

\textbf{Canonical minimum scattering} -- In simple terms, a radiating element is a minimum scattering antenna if it is ``invisible'' when open circuited \cite{RenzoDT22}. Thus, an RIS element whose tunable impedance is set to infinity results in no scattering. This is, of course, an approximation, because the two arms of a dipole always scatter some signal if illuminated by an external source. {\textcolor{black}{The assumption of canonical minimum scattering elements is apparent by direct analysis of ${{\bf{H}}_Z}$ in \eqref{Eq:HZ}}}. We see that the second addend in \eqref{Eq:HZ} tends to zero if the elements of the matrix ${{{\bf{Z}}_{\rm{S}}}}$ tend to infinity. The canonical minimum scattering condition is approximately fulfilled for RIS elements that are sufficiently small to be approximated as ideal radiators \cite{DR1}. The generalization of the multiport network model in \eqref{Eq:HZ} to scattering elements that do not fulfill the condition of canonical minimum scattering is an aspect open to research.

{\textcolor{black}{\textbf{Multiport network theory vs. communication theory}}} -- The multiport network model for RIS-aided communications in \eqref{Eq:HS} and \eqref{Eq:HZ}, despite some modeling assumptions, allows us to unveil some assumptions implicitly made in communication theory (CT). Specifically, the end-to-end channel typically utilized in communication theory can be formulated as follows:
\begin{equation} \label{Eq:HCT}
{\textcolor{black}{{{\bf{H}}_{{\rm{CT}}}} = {{\bf{H}}_{{\rm{RT}}}} + {{\bf{H}}_{{\rm{RS}}}}{{\bf{\Gamma }}_{\rm{H}}}{{\bf{H}}_{{\rm{ST}}}}}}
\end{equation}
\noindent {\textcolor{black}{where, as customary in communication theory, the wireless channels are denoted by ${\bf{H}}$. Let us focus on the RIS-scattered component in  \eqref{Eq:HCT}, i.e., ${{\bf{H}}_{{\rm{RS}}}}{{\bf{\Gamma }}_{\rm{H}}}{{\bf{H}}_{{\rm{ST}}}}$, and assume that the wireless channels ${\bf{H}}$ in \eqref{Eq:HCT} correspond to the scattering matrices ${\bf{S}}$ in \eqref{Eq:HS}. Details about the equivalence ${\bf{H}} \leftrightarrow {\bf{S}}$ are given next. By comparing \eqref{Eq:HCT} with \eqref{Eq:HS}, we evince that the model in \eqref{Eq:HCT} holds true if and only if ${{{\bf{S}}_{{\rm{emc}}}}} = {\bf{0}}$.}} {\textcolor{black}{This implies the following:
\begin{itemize}
\item Since the off-diagonal elements of ${{{\bf{S}}_{{\rm{emc}}}}}$ are zero, ${\bf{H}}_{{\rm{CT}}}$ in \eqref{Eq:HCT} ignores the electromagnetic mutual coupling;
\item Since the diagonal elements of ${{{\bf{S}}_{{\rm{emc}}}}}$ are zero, ${\bf{H}}_{{\rm{CT}}}$ in \eqref{Eq:HCT} can be used if and only if the self-impedances at the ports of the RIS are equal to the reference impedance $Z_0$.
\end{itemize}
}}
\noindent According to ${\bf{H}}_{{\rm{CT}}}$, as a result, an RIS operates as an ideal device, whose scattering properties are uniquely determined by the tunable loads connected to its ports, i.e., by ${{{\bf{\Gamma }}_{\rm{S}}}}$. If the diagonal elements of ${{{\bf{\Gamma }}_{\rm{S}}}}$ have unit modulus, i.e., the tunable impedances of the RIS are imaginary numbers, there are no losses, and the RIS applies a simple phase shift to the impinging electromagnetic waves. The RIS model typically utilized in wireless communications ignores, hence, important aspects that \eqref{Eq:HS} and \eqref{Eq:HZ} inherently consider. This includes the electromagnetic mutual coupling, and the interrelation between the amplitude and phase transformations applied to the incident waves, which is determined by the terms ${\left( {{\bf{I}} - {{\bf{\Gamma }}_{\rm{S}}}{\mathbf{S}_{{\rm{emc}}}}} \right)^{ - 1}}$ and ${{\left( {{\mathbf{Z}_{{\rm{emc}}}} + {{\bf{Z}}_{\rm{S}}}} \right)}^{ - 1}}$ in \eqref{Eq:HS} and \eqref{Eq:HZ}, respectively.

\textbf{Structural scattering} -- {\textcolor{black}{There exist other key aspects that the electromagnetically consistent models in \eqref{Eq:HS} and \eqref{Eq:HZ} account for, but the model in \eqref{Eq:HCT} does not.}} As shown in \cite{li2023diagonal}, the channel models in \eqref{Eq:HS} and \eqref{Eq:HZ} are equivalent to one another. Specifically, the following relations hold true \cite[Eq. (6)]{li2023diagonal}:
\begin{align}
& {{\bf{S}}_{{\rm{RS}}}} = \frac{{{{\bf{Z}}_{{\rm{RS}}}}}}{{2{Z_0}}}{\bf{I}} - \frac{{{{\bf{Z}}_{{\rm{RS}}}}}}{{2{Z_0}}}{\left( {{{\bf{Z}}_{{\rm{SS}}}} + {Z_0}{\bf{I}}} \right)^{ - 1}}\left( {{{\bf{Z}}_{{\rm{SS}}}} - {Z_0}{\bf{I}}} \right) \nonumber \\
& {{\bf{S}}_{{\rm{SS}}}} = {\left( {{{\bf{Z}}_{{\rm{SS}}}} + {Z_0}{\bf{I}}} \right)^{ - 1}}\left( {{{\bf{Z}}_{{\rm{SS}}}} - {Z_0}{\bf{I}}} \right) \nonumber \\
& {{\bf{S}}_{{\rm{ST}}}} = {\left( {{{\bf{Z}}_{{\rm{SS}}}} + {Z_0}{\bf{I}}} \right)^{ - 1}}{{\bf{Z}}_{{\rm{ST}}}} \nonumber \\
& {{\bf{S}}_{{\rm{RT}}}} = \frac{{{{\bf{Z}}_{{\rm{RT}}}}}}{{2{Z_0}}} - \frac{{{{\bf{Z}}_{{\rm{RS}}}}}}{{2{Z_0}}}{\left( {{{\bf{Z}}_{{\rm{SS}}}} + {Z_0}{\bf{I}}} \right)^{ - 1}}{{\bf{Z}}_{{\rm{ST}}}} \label{Eq:StructScatt}
\end{align}
{\textcolor{black}{The communication channels in \eqref{Eq:HS} and \eqref{Eq:HZ} are equivalent in the sense that ${{\bf{H}}_{{\rm{S}}}} = {{\bf{H}}_{{\rm{Z}}}}$. However, the individual terms in \eqref{Eq:HS} and \eqref{Eq:HZ} are not equivalent to one another. A major aspect is the different physical meaning between ${{\bf{S}}_{{\rm{RT}}}}$ in \eqref{Eq:HS} and ${{{\bf{Z}}_{{\rm{RT}}}}}$ in \eqref{Eq:HZ}, especially compared with ${{\bf{H}}_{{\rm{RT}}}}$ in \eqref{Eq:HCT}. Specifically, ${{\bf{H}}_{{\rm{RT}}}}$ in \eqref{Eq:HCT} is usually associated with the direct link between the transmitter and receiver: If this latter link is physically blocked by objects, ${{\bf{H}}_{{\rm{RT}}}} = {\bf{0}}$ is assumed. This is, however, not consistent with ${{\bf{H}}_{{\rm{S}}}}$ in \eqref{Eq:HS} and ${{\bf{H}}_{{\rm{Z}}}}$ in \eqref{Eq:HZ}. First of all, the physical transmitter-receiver direct link corresponds to the impedance matrix ${{{\bf{Z}}_{{\rm{RT}}}}}$ in \eqref{Eq:HZ}. The reason is that the entries of ${{{\bf{Z}}_{{\rm{RT}}}}}$ model the electric field generated by the transmitter and observed at the receiver \cite{DR1}. This is not the case for the scattering matrix ${{{\bf{S}}_{{\rm{RT}}}}}$ in \eqref{Eq:HS}. From \eqref{Eq:StructScatt}, in fact, we see that ${{{\bf{S}}_{{\rm{RT}}}}} \ne {{{\bf{Z}}_{{\rm{RT}}}}}$ and we evince that ${{{\bf{S}}_{{\rm{RT}}}}} \ne {\bf{0}}$, even if the physical direct transmitter-receiver link is blocked by obstacles, i.e., ${{{\bf{Z}}_{{\rm{RT}}}}} = {\bf{0}}$: ${{{\bf{S}}_{{\rm{RT}}}}}$ is given by the summation of the physical direct transmitter-receiver link and another term that includes the antenna structural scattering \cite{abs-2308-16856}. The structural scattering is always present regardless of whether the direct transmitter-receiver link is physically blocked. This aspect is not captured by the channel model utilized in communication theory, i.e., \eqref{Eq:HCT}. The structural scattering from an RIS is considered in ${{\bf{H}}_{{\rm{Z}}}}$ as well, but it is hidden in the second addend in \eqref{Eq:HZ}.}}

\begin{figure}[!t]
\centering
\includegraphics[width=0.90\columnwidth]{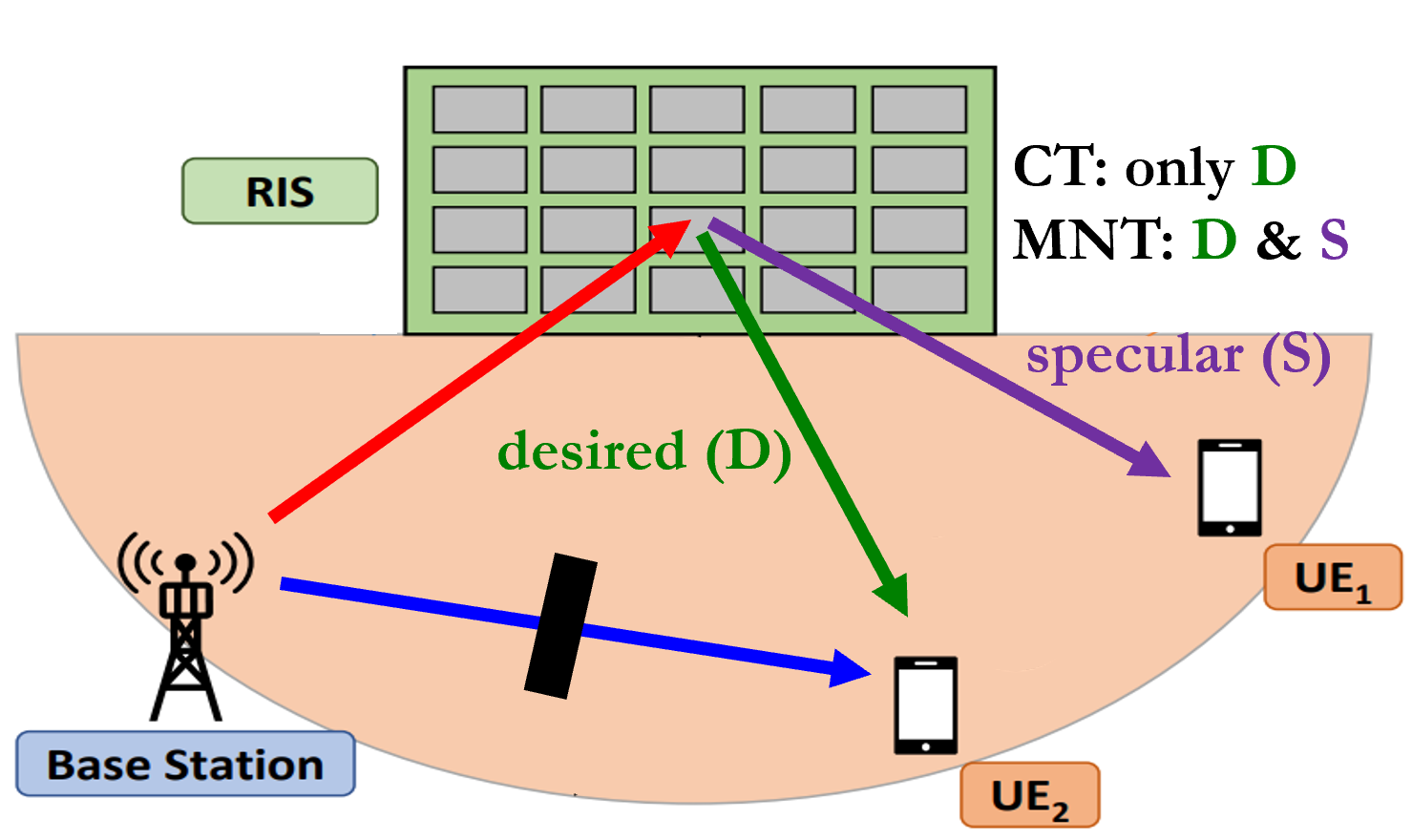}
\caption{{\textcolor{black}{Structural scattering: ${{\bf{H}}_{{\rm{CT}}}}$ (CT) vs.  ${{\bf{H}}_{{\rm{Z}}}}$ and ${{\bf{H}}_{{\rm{S}}}}$ (MNT).}}}
\label{Fig:Fig_MNT_StrSc} \vspace{-0.25cm}
\end{figure}
\textbf{{\textcolor{black}{Physical meaning of the structural scattering: Unwanted re-radiations}}} -- The presence of structural scattering in RIS-aided channels unveils a major difference between the channel model ${{\bf{H}}_{{\rm{CT}}}}$ and the electromagnetically consistent channel models ${{\bf{H}}_{{\rm{S}}}}$ and ${{\bf{H}}_{{\rm{Z}}}}$. From \eqref{Eq:StructScatt}, the structural scattering is
\begin{align} \label{Eq:StSc}
{{\bf{S}}_{{\rm{StSc}}}} =  - \frac{{{{\bf{Z}}_{{\rm{RS}}}}}}{{2{Z_0}}}{\left( {{{\bf{Z}}_{{\rm{SS}}}} + {Z_0}{\bf{I}}} \right)^{ - 1}}{{\bf{Z}}_{{\rm{ST}}}}
\end{align}
Comparing \eqref{Eq:StSc} and \eqref{Eq:HS}, we evince that the structural scattering in \eqref{Eq:StSc} corresponds to the field scattered from the RIS when the tunable impedances are all equal to the reference impedance $Z_0$, i.e., ${{\bf{Z}}_{\rm{S}}} = {Z_0}{\bf{I}}$, which corresponds to ${{{\bf{\Gamma }}_{\rm{S}}}} = {\bf{0}}$ in \eqref{Eq:HS}. In other words, an RIS {\textcolor{black}{re-radiates}} an electromagnetic wave even if its reflection coefficient is equal to zero. This is in stark contrast with ${{\bf{H}}_{{\rm{CT}}}}$ in \eqref{Eq:HCT}, which provides no re-radiated signal if ${{{\bf{\Gamma }}_{\rm{S}}}} = {\bf{0}}$. In a nutshell, ${{\bf{S}}_{{\rm{StSc}}}}$ is always present in the re-radiated signal from an RIS, even if the direct transmitter-receiver link is physical blocked and the reflection coefficient of the RIS is zero. {\textcolor{black}{As illustrated in Fig. \ref{Fig:Fig_MNT_StrSc}, this has major implications when optimizing an RIS: The term ${{\bf{S}}_{{\rm{StSc}}}}$ corresponds to a specular reflection. This specular reflection may interfere with the incident signal for normal incidence or with other users (as shown Fig. \ref{Fig:Fig_MNT_StrSc}), but it is usually ignored in communication theory. This implies that an RIS needs to be optimized for steering the electromagnetic waves towards the desired direction of {\textcolor{black}{re-radiation}}, but also for reducing the structural scattering from it, so as not to reduce the power scattered towards the intended direction and to avoid interference with other users, including the transmitter. This optimization problem has recently been analyzed in \cite{abs-2308-16856}.}}

{\textcolor{black}{\textbf{${{\bf{H}}_{{\rm{CT}}}}$ vs. ${{\bf{H}}_{{\rm{S}}}}$ and ${{\bf{H}}_{{\rm{Z}}}}$} -- In a nutshell, the key differences between the communication theoretic model in \eqref{Eq:HCT} and the multiport network models in \eqref{Eq:HS} and \eqref{Eq:HZ} are the following:
\begin{itemize}
\item The impedance matrices in \eqref{Eq:HZ} model the physical links between pairs of devices. ${{\bf{H}}_{{\rm{Z}}}}$ in \eqref{Eq:HZ} includes the structural scattering from the RIS in ${{\bf{Z}}_{{\rm{RS}}}}{{\left( {{\mathbf{Z}_{{\rm{emc}}}} + {{\bf{Z}}_{\rm{S}}}} \right)}^{ - 1}}{{\bf{Z}}_{{\rm{ST}}}}$.
\item The scattering matrices in \eqref{Eq:HS} do not model, in general, only the physical links between pairs of devices. Equation \eqref{Eq:StructScatt} relates the scattering matrices in \eqref{Eq:HS} with the physical links, i.e., the impedance matrices in \eqref{Eq:HZ}. ${{\bf{H}}_{{\rm{S}}}}$ in \eqref{Eq:HS} includes the structural scattering from the RIS in ${{{\bf{S}}_{{\rm{RT}}}}}$.
\item The wireless channels in \eqref{Eq:HCT} are meant to model the physical links between pairs of devices. However, the formulation in terms of reflection coefficient makes \eqref{Eq:HCT} more similar to \eqref{Eq:HS} than to \eqref{Eq:HZ}. Specifically, ${{\bf{H}}_{{\rm{RT}}}}$ is more physically related to ${{\bf{Z}}_{{\rm{RT}}}}$, while ${{\bf{H}}_{{\rm{ST}}}}$ and ${{\bf{H}}_{{\rm{RS}}}}$ are more mathematically related to ${{\bf{S}}_{{\rm{ST}}}}$ and ${{\bf{S}}_{{\rm{RS}}}}$, respectively. ${{\bf{H}}_{{\rm{CT}}}}$ ignores the structural scattering from the RIS.
\item The multiport network models in ${{\bf{H}}_{{\rm{S}}}}$ in \eqref{Eq:HS} and ${{\bf{H}}_{{\rm{Z}}}}$ in \eqref{Eq:HZ} account for the electromagnetic mutual coupling, and for the interrelation between the amplitude and phase applied to the scattered signal. The channel ${{\bf{H}}_{{\rm{CT}}}}$ in \eqref{Eq:HCT} does not.
\end{itemize}
}}

\begin{figure}[!t]
\centering
\includegraphics[width=0.90\columnwidth]{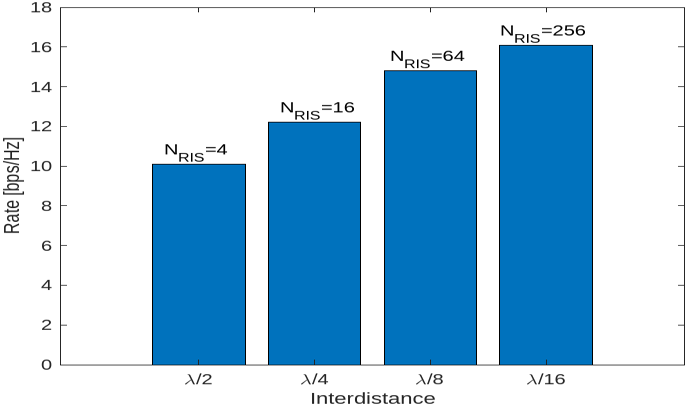}
\caption{{\textcolor{black}{Mutual coupling aware optimization: The size of the RIS is kept fixed and the number of RIS elements is determined by the interdistance.}}}
\label{Fig:Fig_MNT_MC} \vspace{-0.25cm}
\end{figure}
\textbf{Mutual coupling aware optimization} -- The optimization of an RIS based on ${{\bf{H}}_{{\rm{S}}}}$ and ${{\bf{H}}_{{\rm{Z}}}}$ is usually more difficult than the optimization based on ${{\bf{H}}_{{\rm{CT}}}}$. The main reason is that the optimization variables ${{{\bf{\Gamma }}_{\rm{S}}}}$ in \eqref{Eq:HS} and ${{{\bf{Z}}_{\rm{S}}}}$ in \eqref{Eq:HZ}, appear in the inverse of a matrix, which, in the presence of mutual coupling, is a full matrix that is difficult to invert. Thus, new optimization algorithms are needed. This aspect is tackled in \cite{SARIS, DR3} and references therein, where the authors utilize the Neumann series approximation and the Gram-Schmidt orthogonalization to deal with the matrix inversion. The results in \cite{SARIS} and \cite{DR3} show that exploiting the mutual coupling leads to major gains in terms of achievable rate, especially if the size of the RIS is kept fixed and the number of RIS elements is increased, {\textcolor{black}{as shown in Fig. \ref{Fig:Fig_MNT_MC}.}} If the mutual coupling is ignored, the performance degradation may be significant.

\textbf{Modeling the multipath: {\textcolor{black}{Beyond free space}}} -- The multiport network models in \eqref{Eq:HS} and \eqref{Eq:HZ} can be applied in free space. This is the usual reference scenario in electromagnetics. In wireless communications, on the other hand, multipath propagation cannot be ignored. The authors of \cite{SARIS} have recently generalized, in an electromagnetically consistent manner, the impedance model in \eqref{Eq:HZ}, by considering the scattering from natural objects. The approach relies on the discrete dipole approximation from \cite{1236083}: Any natural and engineered scattering object is approximated with a set of thin wire dipoles loaded with impedances that depend on the material properties of the objects. The discrete dipole approximation in \cite{1236083} is especially suited for integration in a multiport network model, since the scattering objects are modeled as multiport networks.

\textbf{{\textcolor{black}{Modeling the multipath: Multiport network theory vs. communication theory}}} -- Embracing the discrete dipole approximation for natural scatterers, the authors of \cite{SARIS} have made a major finding in terms of modeling the multipath. Conventional channel models often adopt an additive multipath model, i.e., given the channel in free space, the multipath is added to it. By contrast, the authors of \cite{SARIS} show that the presence of multipath results in more complex interactions (secondary scattering effects) in the presence of an RIS, which lead to a non-linear and non-additive multipath contribution. The authors of \cite{SARIS} prove that the term ${{{\left( {{{\bf{Z}}_{{\rm{SS}}}} + {{\bf{Z}}_{\rm{S}}}} \right)}^{ - 1}}}$ in \eqref{Eq:HZ} is replaced by the term  ${\left( {{{\bf{Z}}_{{\rm{SS}}}} + {{\bf{Z}}_{\rm{S}}} + {{\bf{Z}}_{{\rm{SOS}}}}} \right)^{ - 1}}$, where the impedance matrix ${{{\bf{Z}}_{{\rm{SOS}}}}}$ models the interactions between the scattering objects in the environment and the RIS. Only if such interactions are negligible, the conventional additive model is retrieved \cite[Sec. II-A]{SARIS}. This corroborates the need of electromagnetically consistent models in wireless communications.

\textbf{Takeaway} -- Using multiport network theory, we have presented an electromagnetically consistent model for wireless communications that accounts for the mutual coupling, and that can be applied to free space and multipath channels. We have specialized the approach to RIS-aided channels, and have shown that conventional scattering models utilized in communication theory overlook fundamental electromagnetic aspects. This includes the amplitude-phase dependence of the scattered electromagnetic field, the structural scattering that results in unwanted {\textcolor{black}{re-radiated}} beams, and the complex interactions between natural and engineered scattering objects.

\section{Spatial Multiplexing in the Near Field}
Near field communication in line-of-sight channels is gaining prominence in wireless systems \cite{an2023beamfocusingaided}, {\textcolor{black}{\cite{10273772}}}. This is due to the large electrical size of emerging antenna technologies, including RIS and HoloS \cite{an2023stacked}, \cite{RenzoZDAYRT20}, and the potential use of high frequency bands for communication, such as the sub-terahertz and terahertz spectra \cite{10133498}. Increasing the size of the transmission and reception devices and/or the carrier frequency, in fact, increases the far-field (Fraunhofer) distance, which is conventionally defined as ${d_{\rm{FF}}={{2D^{2}}/ {\lambda }}}$ when $D$ is the largest dimension of the radiating device. Near field propagation offers new opportunities but presents unique challenges for communication system design \cite{an2023beamfocusingaided}. We focus on one peculiar characteristic of near field communication: Realizing spatial multiplexing in line-of-sight channels, i.e., in the absence of multipath or in the presence of poor multipath \cite{10133498}.

\textbf{Spatial multiplexing in line-of-sight channels} -- In contemporary MIMO communication systems, spatial multiplexing is often implicitly associated with the possibility of transmitting and receiving multiple data streams in an optimal, interference-free manner, over wireless channels that are characterized by rich scattering (multipath channels). In these scenarios, the rank of the MIMO channel matrix, which is equal to the minimum between the number of antennas at the transmitter and the number of antennas at the receiver, corresponds to the spatial multiplexing gain. As recently overviewed in \cite{10133498}, however, spatial multiplexing is possible in line-of-sight channels as well, and it has been a subject of intense study for many years. While spatial multiplexing in multipath channels can be obtained at distances larger than the Fraunhofer distance, spatial multiplexing in line-of-sight channels is possible only in the near field. Thanks to the current interest in emerging electrically large MIMO technologies, e.g., RIS and HoloS, and in exploiting the sub-terahertz and terahertz frequency bands, near field communication in line-of-sight (or poor scattering) channels has become a subject of renewed interest, with focus on understanding the ultimate performance limits and designing the encoding, decoding, and signal processing algorithms for attaining them.

\textbf{Realizing spatial multiplexing in line-of-sight channels} -- {\textcolor{black}{Spatial multiplexing in line-of-sight channels is not a new subject. Here, we overview the best known architectures that support spatial multiplexing in line-of-sight channels \cite{10133498}.}}
\begin{itemize}
\item \textbf{Optimally spaced antennas} -- The first architecture is based on optimally spacing the antenna elements so that the corresponding MIMO line-of-sight channel matrix  has full rank. This solution ensures that the spatial multiplexing gain coincides with the number of available antennas. The downside is, however, that the locations of the antennas need to be optimized as a function of the locations of the transmitter and receiver. Also, the resulting MIMO architecture has a sparse structure with antenna elements that are several wavelengths far apart.
\item \textbf{$\lambda/2$-spaced antennas} -- The second architecture is based on spacing the antenna elements at distances equal to half of the wavelength. This solution originates from Petersen-Middleton's sampling theorem (discussed in Section II) applied to the plane wave spectrum representation restricted to the visible range. This architecture has the advantage of not requiring the optimization of the locations of the antennas as a function of those of the transmitter and receiver. The excess of antenna elements with respect to the architecture with optimally spaced antennas provides a better beamforming gain, but it typically requires a number of radio frequency chains that is larger than the rank of the MIMO channel matrix.
\item \textbf{HoloS} -- The third architecture is the HoloS. In practical terms, a HoloS can be thought of as a hybrid MIMO architecture with a number of radio frequency chains that coincides with the rank of the line-of-sight channel between two communicating HoloSs. This channel is referred to as the holographic MIMO channel. The radio frequency chains feed, however, a continuous surface that radiates the electromagnetic waves. Some signal processing operations are hence performed in the digital domain and some others in the wave domain. A HoloS strikes an appealing tradeoff between the number of radio frequency chains and its capability of finely controlling the radiated and received electromagnetic fields in the wave domain. The number of radio frequency chains needs, however, to be adapted to the rank of the holographic line-of-sight MIMO channel, which highly depends on the network topology (e.g., the location of the transmitter and receiver), and the orientation and tilt of the HoloS.
\end{itemize}

{\textcolor{black}{\textbf{The holographic line-of-sight MIMO channel}}} -- The three mentioned architectures have their own advantages and limitations. The HoloS is an emerging technology that offers the highest design flexibility in terms of signal processing optimization in the digital and wave domains. Compared with the other MIMO architectures, a HoloS is a continuous structure, i.e., a surface not a multiple (discrete) antenna device. This is an important aspect in light of the presence of electromagnetic mutual coupling in MIMO architectures whose antenna elements have interdistances smaller than half of the wavelength (Section II). Due to this major difference between continuous and sparse or half-wavelength spaced antenna elements, the mathematical tools for studying a HoloS are different from those utilized for analyzing conventional MIMO antennas. Also, the concept of \textit{rank of a holographic line-of-sight MIMO channel}, which we have mentioned in the previous paragraph, deserves a proper definition, as compared with the typical definition of \textit{rank of a line-of-sight MIMO channel matrix}, whose definition is well-known and not ambiguous.

\textbf{Tools for MIMO: Singular value decomposition for matrices} -- In a typical MIMO system, the channel between the transmitter and receiver is formulated in a matrix form, i.e., ${\bf{y}} = {\bf{Hx}} + {\bf{n}}$, where $\bf{y}$ is the received signal vector, $\bf{x}$ is the transmitted signal vector, and $\bf{n}$ is the noise vector. The MIMO channel matrix $\bf{H}$ is a rectangular matrix with a number of rows and a number of columns equal to the number of antennas at the receiver and transmitter, respectively. The usual approach for analyzing the spatial multiplexing properties of a MIMO channel is to apply the singular value decomposition (SVD) to $\bf{H}$, and to identify the number of non-zero eigenvalues (the rank of $\bf{H}$). In line-of-sight channels, the entries of $\bf{H}$ are deterministic values that are determined by the geometry of the MIMO channel under consideration, e.g., the positions of the transmitter and receiver, and the orientation of the MIMO arrays. In this case, the interdistances between the antennas in the MIMO arrays are optimized for ensuring the maximum rank. {\textcolor{black}{Consider two MIMO linear antennas with the same number of elements $N$, which are faced against each other at a distance $D$ in the near field region of each other. The optimal interdistance between the antenna elements, which provides a rank equal to $N$, is ${d_{{\rm{opt}}}} = \sqrt {\lambda D/N}$ (Rayleigh criterion) \cite{10133498}. It is apparent that the interdistance ${d_{{\rm{opt}}}}$ may be much larger than the wavelength in practical network deployments.}}

\textbf{Limitations of the SVD for HoloS} -- The SVD is, however, not a suitable approach for analyzing and optimizing HoloSs. First of all, the SVD is a numerical method. Thus, it provides a numerical solution that does not usually offer useful design information. In addition, a HoloS is an antenna structure that has some physical ports, but that it is essentially a continuous surface. The corresponding holographic MIMO channel is, therefore, not well characterized by a channel matrix. In principle, the surface can be discretized, an equivalent MIMO channel matrix can be obtained, and the SVD can be applied. Even if this approach is possible, it does not provide information on the ultimate performance limits of a HoloS, as it is a numerical approach. In addition, the analysis is affected by the discretization applied to the surface, and, more importantly, the discrete elements that constitute the sampled version of the HoloS cannot be interpreted as antenna elements, since the electromagnetic mutual coupling is not explicitly considered. From the numerical point of view, furthermore, the application of the SVD to very large matrices has a high computational complexity, large memory requirements, and is affected by numerical errors and may be potentially be ill-conditioned.

\textbf{Tools for HoloS: Compact and self-adjoint operators} -- A different methodology and different tools are, therefore, needed. To exemplify the approach, we analyze the case study for scalar electromagnetic waves. More information about the vector case can be found in \cite{ruizsicilia2023degrees}. The electric field, $E\left( {{{\bf{r}}_{Rx}}} \right)$, at a receiving HoloS can be expressed as a function of the surface current, ${J\left( {{{\bf{r}}_{Tx}}} \right)}$, at a transmitting HoloS, as follows:
\begin{equation} \label{Eq:Erx}
E\left( {{{\bf{r}}_{Rx}}} \right) = \int_{{{\mathcal{S}}_{Tx}}} {{G_0}\left( {\left| {\bf{r}} \right|} \right)J\left( {{{\bf{r}}_{Tx}}} \right)d{{\bf{r}}_{Tx}}}
\end{equation}
where ${{{\mathcal{S}}_{Tx}}}$ denotes the surface of the transmitting HoloS, ${{{\bf{r}}_{Tx}}}$ is a generic point of the transmitting HoloS, ${{{\bf{r}}_{Rx}}}$ is a generic point of the receiving HoloS, ${\bf{r}} = {{\bf{r}}_{Rx}} - {{\bf{r}}_{Tx}}$, and ${G_0}\left( {\left| {\bf{r}} \right|} \right) = {g_0}\frac{{\exp \left( { - j\kappa \left| {\bf{r}} \right|} \right)}}{{2\lambda \left| {\bf{r}} \right|}}$ is the Green function in free space, where $g_0$ collects some constant terms and $\kappa = 2 \pi / \lambda$. The spatial multiplexing properties of a HoloS-aided communication channel are, therefore, not characterized by a finite-size matrix, but by the (continuous) operator in \eqref{Eq:Erx}. The operator in \eqref{Eq:Erx} is specified by the kernel $G_0(\cdot)$, it is compact, and its adjoint is well defined \cite{ruizsicilia2023degrees}. Thus, the spatial multiplexing gain, and the optimal encoding and decoding waveforms, which correspond to the optimal current distributions at the transmitter and the optimal electric field configurations at the receiver, and are equivalent to the transmit and receive beamforming vectors in conventional MIMO channels, are the solutions of the following two eigenproblems \cite[Lemma 7]{ruizsicilia2023degrees}:
\begin{equation} \label{Eq:EigenTx}
{\mu _m}{\phi _m}\left( {{{\bf{r}}_{Tx}}} \right) = \int_{{{\mathcal{S}}_{Tx}}} {{G_{Tx}}\left( {{{\bf{r}}_{Tx}},{{{\bf{\bar r}}}_{Tx}}} \right){\phi _m}\left( {{{{\bf{\bar r}}}_{Tx}}} \right)d{{{\bf{\bar r}}}_{Tx}}}
\end{equation}
\begin{equation} \label{Eq:EigenRx}
{\mu _m}{\psi _m}\left( {{{\bf{r}}_{Rx}}} \right) = \int_{{{\mathcal{S}}_{Rx}}} {{G_{Rx}}\left( {{{\bf{r}}_{Rx}},{{{\bf{\bar r}}}_{Rx}}} \right){\psi _m}\left( {{{{\bf{\bar r}}}_{Rx}}} \right)d{{{\bf{\bar r}}}_{Rx}}}
\end{equation}
where ${\phi _m}\left( \cdot \right)$ and ${\psi _m}\left(\cdot \right)$ are the $m$th encoding and decoding orthogonal communication waveforms, respectively, and ${\mu _m}$  is the $m$th real and positive eigenvalue of the eigenproblems. Also, the operators ${G_{Tx}}\left( { \cdot , \cdot } \right)$ and ${G_{Rx}}\left( { \cdot , \cdot } \right)$ are compact and self-adjoint, and they are defined as follows:
\begin{equation}
{G_{Tx}}\left( {{{\bf{r}}_{Tx}},{{{\bf{\bar r}}}_{Tx}}} \right) = \hspace{-0.15cm} \int_{{{\mathcal{S}}_{Rx}}} {\hspace{-0.35cm}{G_0^*}\left( {\left| {{{\bf{r}}_{Rx}} - {{\bf{r}}_{Tx}}} \right|} \right){G_0}\left( {\left| {{{\bf{r}}_{Rx}} - {{{\bf{\bar r}}}_{Tx}}} \right|} \right)d{{\bf{r}}_{Rx}}} \nonumber
\end{equation}
\begin{equation}
{G_{Rx}}\left( {{{\bf{r}}_{Rx}},{{{\bf{\bar r}}}_{Rx}}} \right)  = \hspace{-0.15cm} \int_{{{\mathcal{S}}_{Tx}}} {\hspace{-0.35cm}{G_0^*}\left( {\left| {{{\bf{r}}_{Rx}} - {{\bf{r}}_{Tx}}} \right|} \right){G_0}\left( {\left| {{{{\bf{\bar r}}}_{Rx}} - {{\bf{r}}_{Tx}}} \right|} \right)d{{\bf{r}}_{Tx}}}  \nonumber
\end{equation}
where $(^*)$ denotes the complex conjugate. In simple terms, ${\phi _m}\left( \cdot \right)$, ${\psi _m}\left( \cdot \right)$, and ${\mu _m}$ are equivalent to the eigenvectors and eigenvalues obtained by applying the SVD to a MIMO channel matrix. The key difference is that the eigenfunctions ${\phi _m}\left( \cdot \right)$ and ${\psi _m}\left( \cdot \right)$ are continuous functions, and the number of eigenvalues is, in theory, infinite. The eigenfunctions constitute two orthogonal basis functions \cite[Lemma 7]{ruizsicilia2023degrees}.

\textbf{Spectral theorem for compact and self-adjoint operators} -- As a result, any surface current $J(\cdot)$ at the transmitting HoloS can be formulated as a linear combination of the basis functions ${\phi _m}\left( \cdot \right)$, and any electric field $E(\cdot)$ at the receiving HoloS can be formulated as a linear combination of the basis functions ${\psi _m}\left( \cdot \right)$. This is the spectral theorem for compact and self-adjoint operators \cite[Lemmas 3, 7]{ruizsicilia2023degrees}. For any index $m$, the function ${\phi _m}\left( \cdot \right)$ emitted by the transmitting HoloS contributes only to the function ${\psi _m}\left( \cdot \right)$ observed at the receiving HoloS, and the intensity of the connection, i.e., the so-called coupling intensity, between ${\phi _m}\left( \cdot \right)$ and ${\psi _m}\left( \cdot \right)$, is determined by the eigenvalue $\mu_m$. From a MIMO communication system standpoint, each non-zero eigenvalue $\mu_m$ corresponds to a spatial sub-channel, i.e., a communication mode, that can be established between the transmitting and receiving HoloSs via the pair of functions ${\phi _m}\left( \cdot \right)$ and ${\psi _m}\left( \cdot \right)$.

{\textcolor{black}{\textbf{Effective degrees of freedom: Definition and historical perspective}}} -- In theory, as mentioned, the number of eigenvalues of the eigenproblems in \eqref{Eq:EigenTx} and \eqref{Eq:EigenRx} is infinite. In principle, therefore, an infinite number of communication modes can be established between two HoloSs. There exist, however, a  number of effective eigenvalues whose coupling intensity is dominant with respect to the other eigenvalues. The number of dominant eigenvalues is referred to as the ``number of effective degrees of freedom'' (NeDoF). The concept of DoF dates back to the landmark work of Toraldo di Francia, who introduced the concept for an optical image in 1969 \cite{ToraldodiFrancia:69}; Bucci and Franceschetti, who introduced the concept in electromagnetics in 1989 \cite{29386}; Shiu  \textit{et al.}, who introduced the notion of eDoF for a MIMO communication channel in 2000 \cite{ShiuFGK00}; Miller and Piestun, who, in the same year, characterized the NeDoF between two communicating volumes in the paraxial setting \cite{Miller:00}, \cite{Piestun:00}; and Migliore, who clarified in 2006 the connection between Bucci's and Franceschetti's definition of eDoF based on the approximation theory framework and  Shiu's \textit{et al.} definition of eDoF for MIMO communication channels \cite{1589439}. More recent results and an historical survey can be found in \cite{ruizsicilia2023degrees}, \cite{FranceschettiBook}. In \cite{ruizsicilia2023degrees}, the NeDoF is computed between two arbitrary located, oriented, and tilted HoloSs.

{\textcolor{black}{\textbf{\textbf{Kolmogorov's $N$-width and approximation theory}}} -- In mathematical terms, the NeDoF has strong connections with approximation theory and the so-called Kolmogorov's $N$-width \cite[Def. 6, Lemma 7, Def. 8]{ruizsicilia2023degrees}. In simple terms, the NeDoF is the minimum number of basis functions ${\phi _m}\left( \cdot \right)$ and ${\psi _m}\left( \cdot \right)$ that result in an approximation error no greater than a predetermined value $\epsilon$, where the error is computed according to Kolmogorov's $N$-width for subspaces. It is possible to show that Kolmogorov's $N$-width coincides with the square root of the $(N+1)$th largest eigenvalue (i.e., the singular value) of the eigenproblems in \eqref{Eq:EigenTx} and \eqref{Eq:EigenRx} \cite[Lemma 7]{ruizsicilia2023degrees}, \cite{FranceschettiBook}. {\textcolor{black}{In other words, given the tolerable approximation error $\epsilon$, which is often related to the measurement noise when reconstructing a signal, the $(N+1)$th largest eigenvalue of \eqref{Eq:EigenTx} and \eqref{Eq:EigenRx} is smaller than the approximation error $\epsilon$}}. In mathematical terms, this implies that the representations of the surface current $J(\cdot)$ and the electric field $E(\cdot)$ in terms of basis functions ${\phi _m}\left( \cdot \right)$ and ${\psi _m}\left( \cdot \right)$, respectively, can be truncated to the NeDoF, as
\begin{align}
& J\left( {{{\bf{r}}_{Tx}}} \right) = \sum\limits_{m = 1}^{ + \infty } {{a_m}{\phi _m}\left( {{{\bf{r}}_{Tx}}} \right)}  \approx \sum\limits_{m = 1}^N {{a_m}{\phi _m}\left( {{{\bf{r}}_{Tx}}} \right)} \label{Eq:ApproxAccuracy1} \\
& E\left( {{{\bf{r}}_{Rx}}} \right) = \sum\limits_{m = 1}^{ + \infty } {{b_m}{\psi _m}\left( {{{\bf{r}}_{Rx}}} \right)}  \approx \sum\limits_{m = 1}^N {{b_m}{\psi _m}\left( {{{\bf{r}}_{Rx}}} \right)}  \label{Eq:ApproxAccuracy2}
\end{align}
where $N$ denotes the NeDoF, and $a_m$ and $b_m$ are the coefficients of the representation in terms of basis functions.

\textbf{NeDoF in one-dimensional spaces} -- Based on Kolmogorov's definition, the NeDoF is well defined from the mathematical standpoint. A major research problem is, however, the computation of the NeDoF in a closed-form expression. Considering the eigenproblems in \eqref{Eq:EigenTx} and \eqref{Eq:EigenRx}, an accurate estimate for the NeDoF is known in one-dimensional spaces, i.e., when the HoloS reduces to a line or when considering a time-domain signal \cite[Sec. 2.7.3]{FranceschettiBook}. In these cases, the NeDoF can be formulated as follows:
\begin{align} \label{Eq:NeDoF_1}
{\rm{NeDoF}} &= {N_1} + \frac{1}{{{\pi ^2}}}\log \left( {\frac{{1 - \epsilon }}{\epsilon }} \right)\log \left( {\frac{{\pi {N_1}}}{2}} \right) \nonumber\\ & + o\left( {\log \left( {{N_1}} \right)} \right)
\end{align}
{\textcolor{black}{where $0 < \epsilon < 1$ denotes the approximation error (accuracy) requested in \eqref{Eq:ApproxAccuracy1} and \eqref{Eq:ApproxAccuracy2}, according to Kolmogorov's definition. Using the definition in \cite[Eq. (2.132)]{FranceschettiBook}, NeDoF in \eqref{Eq:NeDoF_1} corresponds to the number of eigenvalues in \eqref{Eq:EigenTx} and \eqref{Eq:EigenRx} whose amplitude (normalized to the amplitude of the largest eigenvalue) is greater than or equal to $\epsilon$. Thus, the dimension $N$ of the signal spaces in \eqref{Eq:ApproxAccuracy1} and \eqref{Eq:ApproxAccuracy2} depends only logarithmically on the approximation accuracy $\epsilon$}}. In \eqref{Eq:NeDoF_1}, $N_1$ corresponds, asymptotically as $N_1 \to \infty$, to the number of significant eigenvalues. Assuming that the eigenvalues are ordered in a non-increasing order, in simple terms, those whose index $m$ is such that $m \le N_1$ are nearly equal to one, while those whose index $n$ is such that $m > N_1$ are nearly equal to zero. The transition region between these two sets of eigenvalues is proportional to ${\log \left( {{N_1}} \right)}$ and is usually very sharp for the considered operator. Conventionally, $N_1$ is defined as the NeDoF. $N_1$ can be formulated in a closed-form expression for signals in the time domain and for holographic lines (i.e., the one-dimensional equivalent of a HoloS). The following relations can be obtained \cite[Sec. 2.7.3]{FranceschettiBook}, \cite{10133498}:
\begin{align}
& {N_1} = \frac{{\Omega T}}{\pi } \quad \quad  \quad \quad \quad \quad {\text{time-domain signals}}\\
& {N_1} = \frac{{{L_{Tx}}{L_{Rx}}}}{{\lambda {d_0}}} {\Upsilon _{Tx,Rx}} \quad {\text{linear arrays}}
\end{align}
where $T$ and $\Omega$ are the duration and the frequency bandwidth of the signal, $L_{Tx}$ and $L_{Rx}$ are the lengths of the transmitting and receiving holographic lines, and $d_0$ is the distance between their centers. Also, ${\Upsilon _{Tx,Rx}}$ is a factor that depends on the orientation and alignment of the holographic lines. This is discussed next when considering the bi-dimensional case, i.e., actual HoloSs. We see that the NeDoF increases with the size of the holographic lines and with the carrier frequency. Also, the shorter the transmission distance the larger the NeDoF.

\textbf{NeDoF in multi-dimensional spaces: The case of Landau's eigenvalue problem} -- The one-dimensional case is well known in the literature since Landau's and Pollak's landmark work in 1962 \cite{FranceschettiBook}. The multi-dimensional case is, on the other hand, much less understood. Specifically, the bi-dimensional case is of interest for the analysis of HoloSs. The most relevant result in this context is due to Landau and was published in 1975 \cite{Landau1975}. Landau's eigenvalue problem is summarized in \cite[Lemma 8]{ruizsicilia2023degrees} for application to HoloSs. When considering the eigenproblems in  \eqref{Eq:EigenTx} and \eqref{Eq:EigenRx}, Landau's result can be applied if and only if the Fourier transform, in the wavenumber domain, of the kernels ${G_{Tx}}\left( { \cdot , \cdot } \right)$ and ${G_{Rx}}\left( { \cdot , \cdot } \right)$ has a rectangular shape, i.e., the kernels ${G_{Tx}}\left( { \cdot , \cdot } \right)$ and ${G_{Rx}}\left( { \cdot , \cdot } \right)$ need to be ideal low-pass filters in the wavenumber domain. The authors of \cite{ruizsicilia2023degrees} prove that this is approximately true, provided that the two HoloSs are not too large. Under Landau's assumption, the NeDoF between two HoloSs polarizes asymptotically to the value $N_2$ defined as
\begin{equation} \label{Eq:NeDoF_2}
{\rm{NeDoF}} \to {N_2} = \frac{{{A_{Tx}}{W_G}}}{{4{\pi ^2}}}
\end{equation}
where ${{A_{Tx}}}$ is the area of the transmitting HoloS and ${{W_G}}$ is the spatial bandwidth, in the wavenumber domain, of the kernels ${G_{Tx}}\left( { \cdot , \cdot } \right)$ and ${G_{Rx}}\left( { \cdot , \cdot } \right)$. Similar to the one-dimensional case in \eqref{Eq:NeDoF_1}, the ordered eigenvalues whose index is less than $N_2$ in \eqref{Eq:NeDoF_2} are nearly equal to one, while those whose index is greater than $N_2$ in \eqref{Eq:NeDoF_2} are nearly equal to zero. The transition region between the significant and negligible eigenvalues is sharp in the bi-dimensional case as well. In contrast to the one-dimensional case, however, an estimate of the width of the transition region as a function of the approximation accuracy $\epsilon$ is, to the best of our knowledge, not known. The authors of \cite{ruizsicilia2023degrees} have recently obtained a simple closed-form expression for the spatial bandwidth $W_G$, as follows \cite[Eq. (23)]{ruizsicilia2023degrees}:
\begin{equation} \label{Eq:WG}
{W_G} = 4{\pi ^2}\frac{{{A_{Rx}}}}{{{\lambda ^2}d_0^2}}{\Psi _{Tx,Rx}}
\end{equation}
where ${{A_{Rx}}}$ is the area of the receiving HoloS,  $d_0$ is the distance between the centers of the HoloSs, and ${\Psi _{Tx,Rx}}$ is a factor that depends on the orientation, tilt, and alignment of the HoloSs. We see that the larger the areas of the HoloSs and the smaller the wavelength and the transmission distance, the larger the number of communication modes. The orientation of the HoloSs has, however, a major impact on the NeDoF as well. The case study where the HoloSs are arbitrarily large compared with the transmission distance is analyzed in \cite{ruizsicilia2023degrees}.

\textbf{Communication waveforms: Optimal basis functions based on Kolmogorov's definition} -- Besides estimating the number of significant communication modes, it is important to identify the optimal communication waveforms for data encoding and encoding, i.e., the basis functions ${\phi _m}\left( \cdot \right)$ and ${\psi _m}\left(\cdot \right)$ in \eqref{Eq:EigenTx} and \eqref{Eq:EigenRx}. This problem is even more challenging than the problem of estimating the NeDoF, and only a few results exist in the literature, which provide closed-form (exact or approximate) expressions for the optimal communication waveforms. The one-dimensional case is well-known and was studied by Slepian and Pollak in 1961 \cite{Slepian1961}. In this case, the optimal eigenfunctions are the prolate spheroidal wave functions (PSWF). As far as the bi-dimensional case of interest for the analysis of HoloSs is concerned, Miller computed the optimal waveforms in 2000, under the assumption that the surface are parallel, face each other, and their centers are aligned (paraxial setting) \cite{Miller:00}. In this case, the optimal communication waveforms can be decomposed into the product of two PSWFs. More recently (in 2023), the authors of \cite{ruizsicilia2023degrees} have extended Miller's result to more general network deployments, and have shown when the optimal communication waveforms can be decomposed into the product of two PSWFs. The computation of the eigenfunctions of \eqref{Eq:EigenTx} and \eqref{Eq:EigenRx} in general network settings is, however, an open research problem.

\begin{figure}[!t]
\centering
\includegraphics[width=0.90\columnwidth]{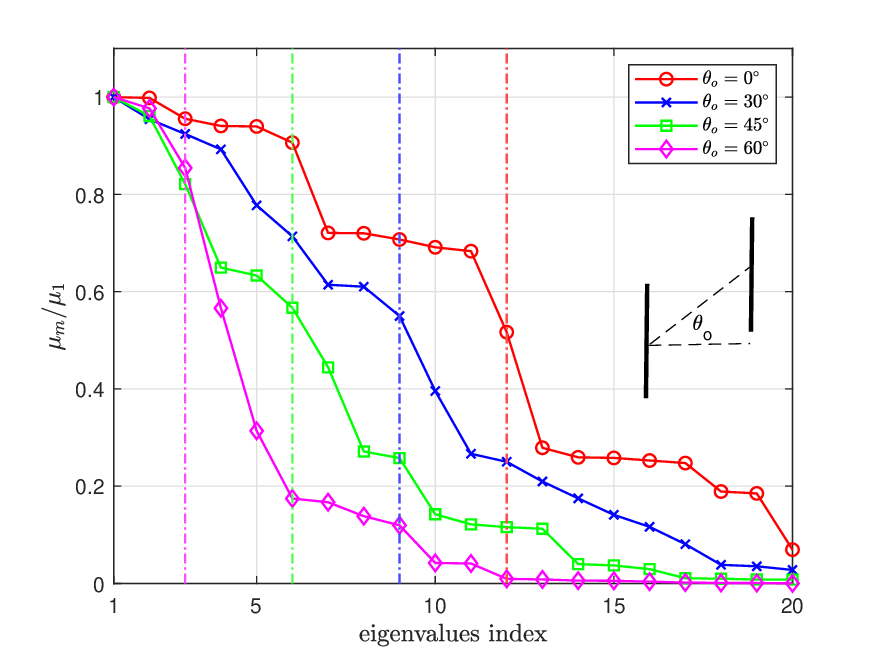}
\caption{Number of effective degrees of freedom. The parameter $\theta_o$ represents the offset between the centers of two HoloSs that face each other: $\theta_o = 0^\circ$ if the centers are perfectly aligned. The larger $\theta_o$, the larger the offset.}
\label{Fig:Fig_NeDoF} \vspace{-0.25cm}
\end{figure}
\textbf{A numerical example: Transmission between two HoloSs} -- To better understand the meaning of the NeDoF, we compare the eigenvalues obtained by solving the eigenproblem in \eqref{Eq:EigenTx}, by using numerical methods, with the estimated NeDoF given by $N_2$ in \eqref{Eq:NeDoF_2} with $W_G$ given by \eqref{Eq:WG}. The results are reported in Fig. \ref{Fig:Fig_NeDoF}, where the vertical lines show the estimate $N_2$. We see, for different setups, that $N_2$ provides a good estimate of the number of eigenvalues whose values are not smaller than half of the value of the largest eigenvalue $\mu_1$. Overall, we evince that more than one communication mode is usually supported by two HoloSs, even in the presence of a relatively large offset between the two surfaces (large values of $\theta_o$).

\textbf{Takeaway} -- A HoloS offers a favorable implementation tradeoff for signal processing in the digital and wave domains. We have shown that HoloS-aided channels can be characterized invoking the theory of compact and self-adjoint operators, and have discussed the notions of eDoF and optimal communication waveforms based on approximation theory. Also, we have presented closed-form expressions for the NeDoF and for the communication waveforms in some relevant case studies.

\section{{\textcolor{black}{The Role of ESIT in Information Theory, Communications, and Signal Processing}}}
{\textcolor{black}{ESIT is an interdisciplinary scientific discipline that offers electromagnetically consistent communication models for the transmission and processing of information. The case studies presented in this paper unveil that ESIT may play an important role in information and communication theories, wireless communications, and signal processing. For example:
\begin{itemize}
\item Accurately modeling and understanding the role of the evanescent waves and near fields may allow us to optimally generating and shaping  the electromagnetic waves;
\item Accurately modeling and understanding the role of the electromagnetic mutual coupling may allow us to evaluate and quantify the differences and similarities between continuous-type and discrete-type antenna-array models;
\item The adoption of electromagnetically consistent models may allow us to design new optimization algorithms that account for the impact of undesired scattering and external interfering sources at the design stage;
\item The adoption of electromagnetically consistent models may allow us to design new and efficient signal processing schemes, e.g., for channel estimation in the presence of mutual coupling and fully-connected control circuits;
\item The adoption of integral operators in communications and signal processing may allow us to identify optimal encoding and decoding schemes that are not known yet.
\end{itemize}
}}

\bibliographystyle{IEEEtran}
\bibliography{biblio}

\begin{thebibliography}{10}
\providecommand{\url}[1]{#1}
\csname url@samestyle\endcsname
\providecommand{\newblock}{\relax}
\providecommand{\bibinfo}[2]{#2}
\providecommand{\BIBentrySTDinterwordspacing}{\spaceskip=0pt\relax}
\providecommand{\BIBentryALTinterwordstretchfactor}{4}
\providecommand{\BIBentryALTinterwordspacing}{\spaceskip=\fontdimen2\font plus
\BIBentryALTinterwordstretchfactor\fontdimen3\font minus
  \fontdimen4\font\relax}
\providecommand{\BIBforeignlanguage}[2]{{%
\expandafter\ifx\csname l@#1\endcsname\relax
\typeout{** WARNING: IEEEtran.bst: No hyphenation pattern has been}%
\typeout{** loaded for the language `#1'. Using the pattern for}%
\typeout{** the default language instead.}%
\else
\language=\csname l@#1\endcsname
\fi
#2}}
\providecommand{\BIBdecl}{\relax}
\BIBdecl

\bibitem{an2023stacked}
J.~An \emph{et~al.}, ``Stacked intelligent metasurface-aided {MIMO} transceiver
  design,'' \emph{arXiv:2311.09814}, 2023.

\bibitem{RenzoZDAYRT20}
M.~{Di Renzo} \emph{et~al.}, ``Smart radio environments empowered by
  reconfigurable intelligent surfaces: How it works, state of research, and the
  road ahead,'' \emph{{IEEE} J. Sel. Areas Commun.}, vol.~38, pp. 2450--2525,
  2020.

\bibitem{LiWR21}
Q.~Li \emph{et~al.}, ``Single-{RF} {MIMO}: {F}rom spatial modulation to
  metasurface-based modulation,'' \emph{{IEEE} Wirel. Commun.}, vol.~28, pp.
  88--95, 2021.

\bibitem{SihlbomPR23}
B.~Sihlbom, M.~I. Poulakis, and M.~{Di Renzo}, ``Reconfigurable intelligent
  surfaces: Performance assessment through a system-level simulator,''
  \emph{{IEEE} Wirel. Commun.}, vol.~30, no.~4, pp. 98--106, 2023.

\bibitem{RenzoDT22}
M.~{Di Renzo} \emph{et~al.}, ``Communication models for reconfigurable
  intelligent surfaces: From surface electromagnetics to wireless networks
  optimization,'' \emph{Proc. {IEEE}}, vol. 110, no.~9, pp. 1164--1209, 2022.

\bibitem{1188558}
D.~Gabor, ``Communication theory and physics,'' \emph{Trans. IRE Professional
  Group Inform. Theory}, vol.~1, no.~1, pp. 48--59, 1953.

\bibitem{4685903}
F.~K. Gruber and E.~A. Marengo, ``New aspects of electromagnetic information
  theory for wireless and antenna systems,'' \emph{IEEE Trans. Antennas
  Propag.}, vol.~56, no.~11, pp. 3470--3484, 2008.

\bibitem{4020419}
T.~K. Sarkar \emph{et~al.}, ``A discussion about some of the
  principles/practices of wireless communication under a {M}axwellian
  framework,'' \emph{IEEE Trans. Antennas Propag.}, vol.~54, no.~12, pp.
  3727--3745, 2006.

\bibitem{4636839}
M.~D. Migliore, ``On electromagnetics and information theory,'' \emph{IEEE
  Trans. Antennas Propag.}, vol.~56, no.~10, pp. 3188--3200, 2008.

\bibitem{6773024}
C.~E. Shannon, ``A mathematical theory of communication,'' \emph{The Bell
  System Technical Journal}, vol.~27, no.~3, pp. 379--423, 1948.

\bibitem{9438650}
M.~D. Migliore, ``The world beneath the physical layer: An introduction to the
  deep physical layer,'' \emph{IEEE Access}, vol.~9, pp. 77\,106--26, 2021.

\bibitem{abs-2308-16856}
A.~Abrardo, A.~Toccafondi, and M.~{Di Renzo}, ``Design of reconfigurable
  intelligent surfaces by using $s$-parameter multiport network theory --
  {O}ptimization and full-wave validation,'' \emph{arXiv:2311.06648}, 2023.

\bibitem{10042168}
F.~Liu, D.-H. Kwon, and S.~Tretyakov, ``Reflectarrays and metasurface
  reflectors as diffraction gratings: A tutorial,'' \emph{IEEE Trans. Propag.
  Mag.}, vol.~65, no.~3, pp. 21--32, 2023.

\bibitem{li2023tunable}
Y.~Li \emph{et~al.}, ``Tunable perfect anomalous reflection using passive
  aperiodic gratings,'' \emph{arXiv:2303.05411}, 2023.

\bibitem{1140193}
E.~Joy and D.~Paris, ``Spatial sampling and filtering in near-field
  measurements,'' \emph{IEEE Trans. Antennas Propag.}, vol.~20, no.~3, 1972.

\bibitem{PETERSEN1962279}
D.~P. Petersen and D.~Middleton, ``Sampling and reconstruction of
  wave-number-limited functions in {$N$}-dimensional euclidean spaces,''
  \emph{Information and Control}, vol.~5, no.~4, pp. 279--323, 1962.

\bibitem{Orfanidis}
S.~J. Orfanidis, \emph{Electromagnetic Waves and Antennas}, 2016.

\bibitem{9798854}
A.~Pizzo \emph{et~al.}, ``Nyquist sampling and degrees of freedom of
  electromagnetic fields,'' \emph{IEEE Trans. Signal Process.}, pp. 3935--3947,
  2022.

\bibitem{HANSEN2021102791}
T.~B. Hansen, ``Array of line sources that produces preselected plane waves,''
  \emph{Wave Motion}, vol. 106, p. 102791, 2021.

\bibitem{8358753}
D.-H. Kwon, ``Lossless scalar metasurfaces for anomalous reflection based on
  efficient surface field optimization,'' \emph{IEEE Antennas Wirel. Propag.
  Lett.}, vol.~17, no.~7, pp. 1149--1152, 2018.

\bibitem{IvrlacN10}
M.~Ivrlac \emph{et~al.}, ``Toward a circuit theory of communication,''
  \emph{{IEEE} Trans. Circuits Syst. - Regul. Pap.}, vol.~57, no.~7, pp.
  1663--1683, 2010.

\bibitem{9838533}
L.~Han \emph{et~al.}, ``Coupling matrix-based beamforming for superdirective
  antenna arrays,'' in \emph{IEEE Int. Conf. Commun.}, 2022, pp. 5159--5164.

\bibitem{li2023diagonal}
H.~Li \emph{et~al.}, ``Beyond diagonal reconfigurable intelligent surfaces with
  mutual coupling: Modeling and optimization,'' \emph{arXiv:2310.02708}, 2023.

\bibitem{DR1}
G.~Gradoni and M.~{Di Renzo}, ``End-to-end mutual coupling aware communication
  model for reconfigurable intelligent surfaces: An electromagnetic-compliant
  approach based on mutual impedances,'' \emph{{IEEE} Wirel. Commun. Lett.},
  vol.~10, no.~5, pp. 938--942, 2021.

\bibitem{1236083}
S.~Tretyakov, S.~Maslovski, and P.~Belov, ``An analytical model of
  metamaterials based on loaded wire dipoles,'' \emph{IEEE Trans. Antennas
  Propag.}, vol.~51, no.~10, pp. 2652--2658, 2003.

\bibitem{akrout2023physically}
M.~Akrout \emph{et~al.}, ``Physically consistent models for intelligent
  reflective surface-assisted communications under mutual coupling and element
  size constraint,'' \emph{arXiv:2302.11130}, 2023.

\bibitem{SARIS}
P.~Mursia \emph{et~al.}, ``{SARIS}: Scattering aware reconfigurable intelligent
  surface model and optimization for complex propagation channels,'' \emph{IEEE
  Wireless Commun. Lett.}, IEEE Early Access, 2023.

\bibitem{DR3}
H.~E. Hassani \emph{et~al.}, ``Optimization of {RIS}-aided {MIMO} - {A}
  mutually coupled loaded wire dipole model,'' \emph{arXiv:2306.09480}, 2023.

\bibitem{an2023beamfocusingaided}
J.~An \emph{et~al.}, ``Toward beamfocusing-aided near-field communications:
  Research advances, potential, and challenges,'' \emph{arXiv:2309.09242},
  2023.

\bibitem{10273772}
P.~Ramezani \emph{et~al.}, ``Exploiting the depth and angular domains for
  massive near-field spatial multiplexing,'' \emph{IEEE Inf. Theory Mag.},
  2023.

\bibitem{10133498}
M.~{Di Renzo}, D.~Dardari, and N.~Decarli, ``{LoS MIMO-arrays vs. LoS
  MIMO-surfaces},'' in \emph{European Conf. Antennas Propag.}, 2023, pp. 1--5.

\bibitem{ruizsicilia2023degrees}
J.~C. Ruiz-Sicilia, M.~{Di Renzo}, M.~D. Migliore, M.~Debbah, and H.~V. Poor,
  ``On the degrees of freedom and eigenfunctions of line-of-sight holographic
  {MIMO} communications,'' \emph{arXiv:2308.08009}, 2023.

\bibitem{ToraldodiFrancia:69}
G.~T. di~Francia, ``Degrees of freedom of an image,'' \emph{J. Opt. Soc. Am.},
  vol.~59, no.~7, pp. 799--804, Jul. 1969.

\bibitem{29386}
O.~M. Bucci and G.~Franceschetti, ``On the degrees of freedom of scattered
  fields,'' \emph{IEEE Trans. Antennas Propag.}, vol.~37, no.~7, 1989.

\bibitem{ShiuFGK00}
D.~Shiu, G.~J. Foschini, M.~J. Gans, and J.~M. Kahn, ``Fading correlation and
  its effect on the capacity of multielement antenna systems,'' \emph{{IEEE}
  Trans. Commun.}, vol.~48, no.~3, pp. 502--513, 2000.

\bibitem{Miller:00}
D.~A.~B. Miller, ``Communicating with waves between volumes: evaluating
  orthogonal spatial channels and limits on coupling strengths,'' \emph{Appl.
  Opt.}, vol.~39, no.~11, pp. 1681--1699, Apr 2000.

\bibitem{Piestun:00}
R.~Piestun and D.~A.~B. Miller, ``Electromagnetic degrees of freedom of an
  optical system,'' \emph{J. Opt. Soc. Am. A}, vol.~17, no.~5, May 2000.

\bibitem{1589439}
M.~D. Migliore, ``On the role of the number of degrees of freedom of the field
  in {MIMO} channels,'' \emph{IEEE Trans. Ant. Propag.}, vol.~54, 2006.

\bibitem{FranceschettiBook}
M.~Franceschetti, \emph{Wave Theory of Information}.\hskip 1em plus 0.5em minus
  0.4em\relax Cambridge Univ., 2017.

\bibitem{Landau1975}
H.~J. Landau, ``On {S}zegö's eingenvalue distribution theorem and
  non-hermitian kernels,'' \emph{J. d{\textquotesingle}Analyse Math.}, no.~1,
  pp. 335--357, 1975.

\bibitem{Slepian1961}
D.~Slepian and H.~O. Pollak, ``Prolate spheroidal wave functions, {F}ourier
  analysis and uncertainty -- {I},'' \emph{Bell Syst. Tech. J.}, vol.~40,
  no.~1, Jan. 1961.

\end{thebibliography}

\end{document}